\documentclass[11pt]{article}
\usepackage{epsfig}
\usepackage{graphics}
\usepackage{longtable}

\begin{document}

\large{\bf Utilizing the Updated Gamma-Ray Bursts and Type Ia Supernovae to \\
\hspace*{1.5cm} Constrain the Cardassian Expansion Model and Dark Energy}

\vspace*{0.5cm}

\centerline{{\bf Jun-Jie Wei$^{1,2}$, Qing-Bo Ma$^{1,2}$ and Xue-Feng Wu$^{1,3,4}$}}

\vspace*{0.5cm}

{\small \noindent $^1$Purple Mountain Observatory, Chinese Academy of Sciences, Nanjing 210008, China; xfwu@pmo.ac.cn\\
$^2$University of Chinese Academy of Sciences, Beijing 100049, China\\
$^3$Chinese Center for Antarctic Astronomy, Nanjing 210008, China\\
$^4$Joint Center for Particle, Nuclear Physics and Cosmology, Nanjing University-Purple Mountain Observatory, Nanjing 210008, China}

\vskip 0.5cm

{\bf Abstract.} We update gamma-ray burst (GRB) luminosity relations among certain spectral and light-curve features with 139 GRBs.
The distance modulus of 82 GRBs at $z>1.4$ can be calibrated with the sample at $z\leq1.4$ by using the cubic spline
interpolation method from the Union2.1 Type Ia supernovae (SNe Ia) set. We investigate the joint constraints on the
Cardassian expansion model and dark energy with 580 Union2.1 SNe Ia sample ($z<1.4$) and 82 calibrated GRBs data ($1.4<z\leq8.2$).
In $\Lambda$CDM, we find that adding 82 high-\emph{z} GRBs to 580 SNe Ia significantly improves
the constrain on $\Omega_{m}-\Omega_{\Lambda}$ plane. In the Cardassian expansion model, the best fit is $\Omega_{m}=
0.24_{-0.15}^{+0.15}$ and $n=0.16_{-0.52}^{+0.30}$ $(1\sigma)$, which is consistent with the $\Lambda$CDM cosmology $(n=0)$
in the $1\sigma$ confidence region. We also discuss two dark energy models in which the equation of state $w(z)$
is parametrized as $w(z)=w_{0}$ and $w(z)=w_{0}+w_{1}z/(1+z)$, respectively. Based on our analysis, we see that our Universe
at higher redshift up to $z=8.2$ is consistent with the concordance model within $1\sigma$ confidence level.

\section{Introduction}

In recent years, the combined observations of nearby and distant Type Ia supernovae (SNe Ia)
have provided strong evidence for the current accelerated expansion of the universe
\cite{Perlmutter98,Schmidt98,Riess98}.
The cause of the acceleration remains unknown. Many authors suggest that the composition of
the Universe may consist of an extra component called dark energy, which may explain the
acceleration of the Universe at the current epoch. For example, the dark energy model with
a constant equation of state $P/\rho\equiv w=-1$ is one of the several possible explanations for the
acceleration. While other models suggest that dark energy changes with time, and there are
many ways to characterize the time variation of dark-energy. Here, we adopt a simple model
in which the dark-energy equation of state can be parameterized by
$P/\rho\equiv w(z)=w_{0}+w_{1}z/(1+z)=w_{0}+w_{1}(1-a)$ \cite{Chevallier01,Linder03}.
where $w_{0}$ is constant, $w_{1}$ represents the time dependence of dark energy, and
$a=1/(1+z)$ is the scale factor. In addition, models where general relativity is modified
can also drive universe acceleration, such as the Cardassian expansion model is a possible
alternative for explaining the acceleration of the universe that invokes no vacuum energy
\cite{Freese02}.

SNe Ia have been considered a perfect standard candle to measure the geometry and dynamics
of the Universe. Unfortunately, the farthest SNe Ia detected so far is only at $z=1.914$
\cite{Jones13}. It is difficult to observe SNe at $z>2$, even with excellent
space-based platforms such as SNAP \cite{Sholl04}. And this is quite limiting because
much of the most interesting evolution of the Universe occurred well before this epoch.
Gamma-ray bursts (GRBs) are the most luminous transient events at cosmological distances.
Owing to their high luminosities, GRBs can be detected out to very high redshifts
\cite{Lamb00}. In fact, the farthest burst detected so far is
GRB 090423, which is at $z=8.2$ \cite{Tanvir09}. \footnote{A photometric redshift of 9.4
for GRB 090429B was reported by \cite{Cucchiara11}.} Moreover, in contrast to SNe Ia,
gamma-ray photons from GRBs are almost immune to dust extinction, so the observed gamma-ray
flux is a direct measurement of the prompt emission energy. Hence, GRBs are potentially
a more promising standard candles than SNe Ia at higher redshifts. The possible use of GRBs
as cosmological probes started to become reality after some empirical luminosity relations
were discovered. These GRB luminosity relations have been proposed as distance indicators, such
as the correlations $\tau_{\rm lag}-L$ \cite{Norris00}, $V-L$ \cite{Fenimore00},
$E_{\rm p}-E_{\rm iso}$ \cite{Amati02}, $E_{\rm p}-L$ \cite{Wei03,Yonetoku04},
$E_{\rm p}-E_{\rm \gamma}$ \cite{Ghirlanda04a}, $\tau_{\rm RT}-L$ \cite{Schaefer07},
and so on. Here the time lag ($\tau_{\rm lag}$) is the time shift between the hard and soft
light curves; the luminosity ($L$) is the isotropic peak luminosity of a GRB; the variability
($V$) of a burst denotes whether its light curve is spiky or smooth, and $V$ can be obtained
by calculating the normalized variance of an observed light curve around a smoothed version
of that light curve \cite{Fenimore00}; ($E_{\rm p}$) is the burst frame peak energy
in the GRB spectrum; ($E_{\rm iso}$) is the isotropic equivalent gamma-ray energy; ($E_{\rm \gamma}$)
is the collimation-corrected gamma-ray energy; and the minimum rise time ($\tau_{\rm RT}$) in
the gamma-ray light curve is the shortest time over which the light curve rises by half of
the peak flux of the pulse. However, Ref.~\cite{Wang11} found that the updated $V-L$ correlation
was quite scattered. Its intrinsic scatter has been larger than the one that could be
expected of a linear relation.

Generally speaking, with these luminosity indicators, one can make use of them as standard
candles for cosmological research. For example, Ref.~\cite{Schaefer03b} constructed the first GRB
Hubble diagram based on nine GRBs using two GRB luminosity indicators. With the $E_{\rm p}
-E_{\rm \gamma}$ relation, Ref.~\cite{Dai04} placed tight constraints on cosmological parameters and
dark energy. Ref.~\cite{Liang05} used a model-independent multivariable GRB luminosity indicator
to constrain cosmological parameters and the transition redshift. Ref.~\cite{Schaefer07} made use
of five luminosity indicators calibrated with 69 events by assuming two adopted cosmological models
to construct the GRB Hubble diagram. Ref.~\cite{Kodama08} suggested that the time variation of
the dark energy is small or zero up to $z\sim6$ using the $E_{\rm p}-L$ relation. Ref.~\cite{Tsutsui09}
extended the Hubble diagram up to $z=5.6$ using 63 gamma-ray bursts (GRBs) via $E_{\rm p}-L$ relation
and found that these GRB data were consistent with the concordance model within $2\sigma$ level.
In a word, a lot of other works in this so-called GRB
cosmology field have been published (please see \cite{Wang11} and \cite{Wei13} for reviews).
However, there is a so-called circularity problem in the calibration of these luminosity relations.
Because of the current poor information on low-$z$ GRBs, these luminosity relations necessarily
depend on the assumed cosmology. Some authors attempted to circumvent the circularity problem
by using a less model-dependent approach, such as the scatter method \cite{Ghirlanda04b,Ghirlanda06},
the luminosity distance method \cite{Liang06}, the Bayesian method \cite{Firmani05,Wang06}, and the method
by fitting relation parameters of GRBs and cosmological parameters simultaneously \cite{Amati08,Li08}.
However, these statistical approaches still can not avoid the circularity problem completely,
because a particular cosmology model is required in doing the joint fitting. This means that
the parameters of the calibrated relations are still coupled to the cosmological
parameters derived from a given cosmological model.

To solve the circularity problem completely, one should calibrate the GRB relations in a cosmology
independent way. Recently, a new method to calibrate GRBs in a cosmological model-independent way
has been presented \cite{Liang08,Wei09,Wei10}. This method is very similar
to the calibration for SNe Ia by measuring Cepheid variables in the same galaxy, and it is free
from the circularity problem. Cepheid variables have been regarded as the first order standard
candles for calibrating SNe Ia which are the secondary standard candles. Similarly, if we regard
SNe Ia as the first order standard candles, we can also calibrate GRBs relations with a large
number of SNe Ia since objects at the same redshift should have the same luminosity distance in
any cosmology. This method is one of the interpolation procedures which obtain the distance moduli
of GRBs in the redshift range of SNe Ia by interpolating from SNe Ia data in the Hubble diagram.
Then, if we assume that the GRB luminosity relations do not evolve with reshift, we can extend
the calibrated luminosity relations to high-\emph{z} and derive the distance moduli of high-\emph{z}
GRBs. From these obtained distance modulus, we can constrain the cosmological parameters.

In this paper, we will try to determine the cosmological parameters and dark energy using both the
updated 139 GRBs and 580 SNe Ia. In Section~2, we will describe the data we will use and our method
of calibration. To avoid any assumption on cosmological models, we will use the distance moduli of
580 SNe Ia from the Union2.1 sample to calibrate five GRB luminosity relations in the redshift range
of SNe Ia sample $(z<1.4)$. Then, the distance moduli of 82 high-\emph{z} GRBs $(z>1.4)$ can be
obtained from the five calibrated GRB luminosity relations. The joint constraints on the Cardassian
expansion model and dark energy with 580 SNe and 82 calibrated GRBs data whose $z>1.4$ will be presented
in Section~3. Finally, we will summarize our findings and present a brief discussion.

\section{Calibrating the updated luminosity relations of GRBs}

\subsection{Observational data and methodology}
As mentioned above, we calibrate the updated luminosity relations of GRBs using low-\emph{z}
events whose distance moduli can be obtained by those of Type Ia supernovae. Actually, we use
the cosmology-independent calibration method developed by Refs.~\cite{Liang08,Wei09,Wei10}.
This method is one of the interpolation procedures which use the
abundant SNe Ia sample to interpolate the distance moduli of GRBs in the redshift range of
SNe Ia sample ($z<1.4$). More recently, the Supernova Cosmology Project collaboration released
their latest SNe Ia dataset known as the Union2.1 sample, which contains 580 SNe detections \cite{Suzuki12}.
Obviously, there are rich SNe Ia data points, and we can make a better interpolation by using this dataset.

Our updated GRB sample includes 139 GRBs wih redshift measurements, there are 57 GRBs
at $z<1.4$ and 82 GRBs at $z>1.4$. This sample is shown in Table~\ref{GRB},
which includes the following information for each GRB: (1) its name; (2) the redshift;
(3) the bolometric peak flux $P_{\rm bolo}$; (4) the bolometric fluence $S_{\rm bolo}$;
(5) the beaming factor $f_{\rm beam}$; (6) the time lag $\tau_{\rm lag}$; (7) the spectral
peak energy $E_{\rm p}$; and (8) the minimum rise time $\tau_{\rm RT}$. All of these data
were obtained from previously published studies. Before GRB 060607, we take all the data directly
from Ref.~\cite{Schaefer07}. We adopt the data between GRB 060707 and GRB 080721 from
Ref.~\cite{Wang11}. For those GRBs detected after July 7th, 2008, we adopt the data
directly from Ref.~\cite{Yonetoku10}. Applying the interpolation method, we can derive the distance
moduli of 57 low-\emph{z} GRBs and calibrate five GRB luminosity relations with this low-\emph{z}
sample, i.e., the $\tau_{\rm lag}-L$ relation, the $E_{\rm p}-L$ relation, the $E_{\rm p}-
E_{\gamma}$ relation, the $\tau_{\rm RT}-L$ relation, and the $E_{\rm p}-E_{\rm iso}$ relation.
The isotropic peak luminosity of a burst is calculated by
\begin{equation}
L=4\pi D_{\rm L}^{2}P_{\rm bolo},
\label{luminosity}
\end{equation}
the isotropic equivalent gamma-ray energy is given by
\begin{equation}
E_{\rm iso}=4\pi D_{\rm L}^{2}S_{\rm bolo}(1+z)^{-1},
\label{energy}
\end{equation}
and the collimation-corrected energy is
\begin{equation}
E_{\rm \gamma}=E_{\rm iso}f_{\rm beam}=4\pi D_{\rm L}^{2}S_{\rm bolo}f_{\rm beam}(1+z)^{-1}.
\label{correctenergy}
\end{equation}
Here, $D_{\rm L}$ is the luminosity distance of the burst, $P_{\rm bolo}$ and $S_{\rm bolo}$
are the bolometric peak flux and fluence of gamma-rays, respectively, while $f_{\rm beam}=
(1-\cos\theta_{\rm jet})$ is the beaming factor, $\theta_{\rm jet}$ is the jet half-opening angle.
We assume each GRB has bipolar jets, and $E_{\rm \gamma}$ is the true energy of the bipolar jets.

For convenience, the luminosity relations involved in this paper can be generally written
in the power-law forms
\begin{equation}
\log y=a+b\log x,
\label{form}
\end{equation}
where $a$ and $b$ are the intercept and slope of the relation, respectively;
$y$ is the luminosity ($L$ in units of erg $\rm s^{-1}$) or energy ($E_{\rm iso}$ or
$E_{\rm \gamma}$ in units of erg); $x$ is the GRB parameters measured in the
rest frame, e.g., $\tau_{\rm lag}(1+z)^{-1}/$(0.1 s), $E_{\rm p}(1+z)/$(300 keV),
$E_{\rm p}(1+z)/$(300 keV), $\tau_{\rm RT}(1+z)^{-1}/$(0.1 s), $E_{\rm p}(1+z)/$(300 keV),
for the 5 two-variable relations above.

\begin{center}
\begin{tiny}
\begin{longtable}{@{}lcccccccccc@{}}
\caption[Luminosities and luminosity indicators.]{Luminosities and luminosity indicators.} \label{GRB} \\
\hline
GRB &  \emph{z}      & $P_{\rm bolo}$ & $S_{\rm bolo}$  & $f_{\rm beam}$  & $\tau_{\rm lag}$ & $E_{\rm p}$ & $\tau_{\rm RT}$              \\
 & &(erg $\rm cm^{-2}$ $\rm s^{-1}$) &(erg $\rm cm^{-2}$)&   & (s) & (keV) & (s)  \\
\hline
970228	&	0.70	&	7.30E-06	$\pm$	4.30E-07	&	$\cdots$	&	$\cdots$	&	$\cdots$	&	$	115	\pm_{	38	}^{	 38	}$	&	0.26	$\pm$	0.04	\\
970508	&	0.84	&	3.30E-06	$\pm$	3.30E-07	&	8.09E-06	$\pm$	8.10E-07	&	0.0795	$\pm$	0.0204	&	0.50	 $\pm$	0.30	&	$	389	\pm_{	40	}^{	40	}$	&	0.71	$\pm$	0.06	 \\
970828	&	0.96	&	1.00E-05	$\pm$	1.10E-06	&	1.23E-04	$\pm$	1.20E-05	&	0.0053	$\pm$	0.0014	&	 $\cdots$	&	$	298	\pm_{	30	}^{	30	}$	&	0.26	$\pm$	0.07	\\
971214	&	3.42	&	7.50E-07	$\pm$	2.40E-08	&	$\cdots$	&	$\cdots$	&	0.03	$\pm$	0.03	&	$	190	 \pm_{	20	}^{	20	}$	&	0.05	$\pm$	0.02	\\
980613	&	1.10	&	3.00E-07	$\pm$	8.30E-08	&	$\cdots$	&	$\cdots$	&	$\cdots$	&	$	92	\pm_{	42	}^{	 42	}$	&	$\cdots$	\\
980703	&	0.97	&	1.20E-06	$\pm$	3.60E-08	&	2.83E-05	$\pm$	2.90E-06	&	0.0184	$\pm$	0.0027	&	0.40	 $\pm$	0.10	&	$	254	\pm_{	25	}^{	25	}$	&	3.60	$\pm$	0.50	 \\
990123	&	1.61	&	1.30E-05	$\pm$	5.00E-07	&	3.11E-04	$\pm$	3.10E-05	&	0.0024	$\pm$	0.0007	&	0.16	 $\pm$	0.03	&	$	604	\pm_{	60	}^{	60	}$	&	$\cdots$	\\
990506	&	1.31	&	1.10E-05	$\pm$	1.50E-07	&	$\cdots$	&	$\cdots$	&	0.04	$\pm$	0.02	&	$	283	 \pm_{	30	}^{	30	}$	&	0.17	$\pm$	0.03	\\
990510	&	1.62	&	3.30E-06	$\pm$	1.20E-07	&	2.85E-05	$\pm$	2.90E-06	&	0.0021	$\pm$	0.0003	&	0.03	 $\pm$	0.01	&	$	126	\pm_{	10	}^{	10	}$	&	0.14	$\pm$	0.02	 \\
990705	&	0.84	&	6.60E-06	$\pm$	2.60E-07	&	1.34E-04	$\pm$	1.50E-05	&	0.0035	$\pm$	0.0010	&	 $\cdots$	&	$	189	\pm_{	15	}^{	15	}$	&	0.05	$\pm$	0.02	\\
990712	&	0.43	&	3.50E-06	$\pm$	2.90E-07	&	1.19E-05	$\pm$	6.20E-07	&	0.0136	$\pm$	0.0034	&	 $\cdots$	&	$	65	\pm_{	10	}^{	10	}$	&	$\cdots$	\\
991208	&	0.71	&	2.10E-05	$\pm$	2.10E-06	&	$\cdots$	&	$\cdots$	&	$\cdots$	&	$	190	\pm_{	20	}^{	 20	}$	&	0.32	$\pm$	0.04	\\
991216	&	1.02	&	4.10E-05	$\pm$	3.80E-07	&	2.48E-04	$\pm$	2.50E-05	&	0.0030	$\pm$	0.0009	&	0.03	 $\pm$	0.01	&	$	318	\pm_{	30	}^{	30	}$	&	0.08	$\pm$	0.02	 \\
000131	&	4.50	&	7.30E-07	$\pm$	8.30E-08	&	$\cdots$	&	$\cdots$	&	$\cdots$	&	$	163	\pm_{	13	}^{	 13	}$	&	0.12	$\pm$	0.06	\\
000210	&	0.85	&	2.00E-05	$\pm$	2.10E-06	&	$\cdots$	&	$\cdots$	&	$\cdots$	&	$	408	\pm_{	14	}^{	 14	}$	&	0.38	$\pm$	0.06	\\
000911	&	1.06	&	1.90E-05	$\pm$	1.90E-06	&	$\cdots$	&	$\cdots$	&	$\cdots$	&	$	986	\pm_{	100	}^{	 100	}$	&	0.05	$\pm$	0.02	\\
000926	&	2.07	&	2.90E-06	$\pm$	2.90E-07	&	$\cdots$	&	$\cdots$	&	$\cdots$	&	$	100	\pm_{	7	}^{	 7	}$	&	0.05	$\pm$	0.03	\\
010222	&	1.48	&	2.30E-05	$\pm$	7.20E-07	&	2.45E-04	$\pm$	9.10E-06	&	0.0014	$\pm$	0.0001	&	 $\cdots$	&	$	309	\pm_{	12	}^{	12	}$	&	0.12	$\pm$	0.03	\\
010921	&	0.45	&	1.80E-06	$\pm$	1.60E-07	&	$\cdots$	&	$\cdots$	&	0.90	$\pm$	0.30	&	$	89	 \pm_{	21.8	}^{	13.8	}$	&	3.90	$\pm$	0.50	\\
011211	&	2.14	&	9.20E-08	$\pm$	9.30E-09	&	9.20E-06	$\pm$	9.50E-07	&	0.0044	$\pm$	0.0011	&	 $\cdots$	&	$	59	\pm_{	8	}^{	8	}$	&	$\cdots$	\\
020124	&	3.20	&	6.10E-07	$\pm$	1.00E-07	&	1.14E-05	$\pm$	1.10E-06	&	0.0039	$\pm$	0.0010	&	0.08	 $\pm$	0.05	&	$	87	\pm_{	18	}^{	12	}$	&	0.25	$\pm$	0.05	 \\
020405	&	0.70	&	7.40E-06	$\pm$	3.10E-07	&	1.10E-04	$\pm$	2.10E-06	&	0.0060	$\pm$	0.0020	&	 $\cdots$	&	$	364	\pm_{	90	}^{	90	}$	&	0.45	$\pm$	0.08	\\
020813	&	1.25	&	3.80E-06	$\pm$	2.60E-07	&	1.59E-04	$\pm$	2.90E-06	&	0.0012	$\pm$	0.0003	&	0.16	 $\pm$	0.04	&	$	142	\pm_{	14	}^{	13	}$	&	0.82	$\pm$	0.10	 \\
020903	&	0.25	&	3.40E-08	$\pm$	8.80E-09	&	$\cdots$	&	$\cdots$	&	$\cdots$	&	$	2.6	\pm_{	1.4	}^{	 0.8	}$	&	$\cdots$	\\
021004	&	2.32	&	2.30E-07	$\pm$	5.50E-08	&	3.61E-06	$\pm$	8.60E-07	&	0.0109	$\pm$	0.0027	&	0.60	 $\pm$	0.40	&	$	80	\pm_{	53	}^{	22	}$	&	0.35	$\pm$	0.15	 \\
021211	&	1.01	&	2.30E-06	$\pm$	1.70E-07	&	$\cdots$	&	$\cdots$	&	0.32	$\pm$	0.04	&	$	46	 \pm_{	8	}^{	6	}$	&	0.33	$\pm$	0.05	\\
030115	&	2.50	&	3.20E-07	$\pm$	5.10E-08	&	$\cdots$	&	$\cdots$	&	0.40	$\pm$	0.20	&	$	83	 \pm_{	53	}^{	22	}$	&	1.47	$\pm$	0.50	\\
030226	&	1.98	&	2.60E-07	$\pm$	4.70E-08	&	8.33E-06	$\pm$	9.80E-07	&	0.0034	$\pm$	0.0008	&	0.30	 $\pm$	0.30	&	$	97	\pm_{	27	}^{	17	}$	&	0.70	$\pm$	0.20	 \\
030323	&	3.37	&	1.20E-07	$\pm$	6.00E-08	&	$\cdots$	&	$\cdots$	&	$\cdots$	&	$	44	\pm_{	90	}^{	 26	}$	&	1.00	$\pm$	0.50	\\
030328	&	1.52	&	1.60E-06	$\pm$	1.10E-07	&	6.14E-05	$\pm$	2.40E-06	&	0.0020	$\pm$	0.0005	&	0.20	 $\pm$	0.20	&	$	126	\pm_{	14	}^{	14	}$	&	$\cdots$	\\
030329	&	0.17	&	2.00E-05	$\pm$	1.00E-06	&	2.31E-04	$\pm$	2.00E-06	&	0.0049	$\pm$	0.0009	&	0.14	 $\pm$	0.04	&	$	68	\pm_{	2.3	}^{	2.2	}$	&	0.66	$\pm$	0.08	 \\
030429	&	2.66	&	2.00E-07	$\pm$	5.40E-08	&	1.13E-06	$\pm$	1.90E-07	&	0.0060	$\pm$	0.0029	&	 $\cdots$	&	$	35	\pm_{	12	}^{	8	}$	&	0.90	$\pm$	0.20	\\
030528	&	0.78	&	1.60E-07	$\pm$	3.20E-08	&	$\cdots$	&	$\cdots$	&	12.50	$\pm$	0.50	&	$	32	 \pm_{	4.7	}^{	5	}$	&	0.77	$\pm$	0.20	\\
040924	&	0.86	&	2.60E-06	$\pm$	2.80E-07	&	$\cdots$	&	$\cdots$	&	0.30	$\pm$	0.04	&	$	67	 \pm_{	6	}^{	6	}$	&	0.17	$\pm$	0.02	\\
041006	&	0.71	&	2.50E-06	$\pm$	1.40E-07	&	1.75E-05	$\pm$	1.80E-06	&	0.0012	$\pm$	0.0003	&	 $\cdots$	&	$	63	\pm_{	13	}^{	13	}$	&	0.65	$\pm$	0.16	\\
050126	&	1.29	&	1.10E-07	$\pm$	1.30E-08	&	$\cdots$	&	$\cdots$	&	2.10	$\pm$	0.30	&	$	47	 \pm_{	23	}^{	8	}$	&	3.90	$\pm$	0.80	\\
050318	&	1.44	&	5.20E-07	$\pm$	6.30E-08	&	3.46E-06	$\pm$	3.50E-07	&	0.0020	$\pm$	0.0006	&	 $\cdots$	&	$	47	\pm_{	15	}^{	8	}$	&	0.38	$\pm$	0.05	\\
050319	&	3.24	&	2.30E-07	$\pm$	3.60E-08	&	$\cdots$	&	$\cdots$	&	$\cdots$	&	$\cdots$	&	0.19	 $\pm$	0.04	\\
050401	&	2.90	&	2.10E-06	$\pm$	2.20E-07	&	$\cdots$	&	$\cdots$	&	0.10	$\pm$	0.06	&	$	118	 \pm_{	18	}^{	18	}$	&	0.03	$\pm$	0.01	\\
050406	&	2.44	&	4.20E-08	$\pm$	1.10E-08	&	$\cdots$	&	$\cdots$	&	0.64	$\pm$	0.40	&	$	25	 \pm_{	35	}^{	13	}$	&	0.50	$\pm$	0.30	\\
050408	&	1.24	&	1.10E-06	$\pm$	2.10E-07	&	$\cdots$	&	$\cdots$	&	0.25	$\pm$	0.10	&	$\cdots$	 &	0.25	$\pm$	0.08	\\
050416	&	0.65	&	5.30E-07	$\pm$	8.50E-08	&	$\cdots$	&	$\cdots$	&	$\cdots$	&	$	15	\pm_{	2.3	}^{	 2.7	}$	&	0.51	$\pm$	0.30	\\
050502	&	3.79	&	4.30E-07	$\pm$	1.20E-07	&	$\cdots$	&	$\cdots$	&	0.20	$\pm$	0.20	&	$	93	 \pm_{	55	}^{	35	}$	&	0.40	$\pm$	0.20	\\
050505	&	4.27	&	3.20E-07	$\pm$	5.40E-08	&	6.20E-06	$\pm$	8.50E-07	&	0.0014	$\pm$	0.0007	&	 $\cdots$	&	$	70	\pm_{	140	}^{	24	}$	&	0.40	$\pm$	0.15	\\
050525	&	0.61	&	5.20E-06	$\pm$	7.20E-08	&	2.59E-05	$\pm$	1.30E-06	&	0.0025	$\pm$	0.0010	&	0.11	 $\pm$	0.02	&	$	81	\pm_{	1.4	}^{	1.4	}$	&	0.32	$\pm$	0.03	 \\
050603	&	2.82	&	9.70E-06	$\pm$	6.00E-07	&	$\cdots$	&	$\cdots$	&	0.03	$\pm$	0.03	&	$	344	 \pm_{	52	}^{	52	}$	&	0.17	$\pm$	0.02	\\
050802	&	1.71	&	5.00E-07	$\pm$	7.30E-08	&	$\cdots$	&	$\cdots$	&	$\cdots$	&	$\cdots$	&	0.80	 $\pm$	0.20	\\
050820	&	2.61	&	3.30E-07	$\pm$	5.20E-08	&	$\cdots$	&	$\cdots$	&	0.70	$\pm$	0.30	&	$	246	 \pm_{	76	}^{	40	}$	&	2.00	$\pm$	0.50	\\
050824	&	0.83	&	9.30E-08	$\pm$	3.80E-08	&	$\cdots$	&	$\cdots$	&	$\cdots$	&	$\cdots$	&	11.00	 $\pm$	2.00	\\
050904	&	6.29	&	2.50E-07	$\pm$	3.50E-08	&	2.00E-05	$\pm$	2.00E-06	&	0.0097	$\pm$	0.0024	&	 $\cdots$	&	$	436	\pm_{	200	}^{	90	}$	&	0.60	$\pm$	0.20	\\
050908	&	3.35	&	9.80E-08	$\pm$	1.50E-08	&	$\cdots$	&	$\cdots$	&	$\cdots$	&	$	41	\pm_{	9	}^{	 5	}$	&	1.50	$\pm$	0.30	\\
050922	&	2.20	&	2.00E-06	$\pm$	7.30E-08	&	$\cdots$	&	$\cdots$	&	0.06	$\pm$	0.02	&	$	198	 \pm_{	38	}^{	22	}$	&	0.13	$\pm$	0.02	\\
051022	&	0.80	&	1.10E-05	$\pm$	8.70E-07	&	3.40E-04	$\pm$	1.20E-05	&	0.0029	$\pm$	0.0001	&	 $\cdots$	&	$	510	\pm_{	22	}^{	20	}$	&	0.19	$\pm$	0.04	\\
051109	&	2.35	&	7.80E-07	$\pm$	9.70E-08	&	$\cdots$	&	$\cdots$	&	$\cdots$	&	$	161	\pm_{	130	}^{	 35	}$	&	1.30	$\pm$	0.40	\\
051111	&	1.55	&	3.90E-07	$\pm$	5.80E-08	&	$\cdots$	&	$\cdots$	&	1.02	$\pm$	0.10	&	$\cdots$	 &	3.20	$\pm$	1.00	\\
060108	&	2.03	&	1.10E-07	$\pm$	1.10E-07	&	$\cdots$	&	$\cdots$	&	$\cdots$	&	$	65	\pm_{	600	}^{	 10	}$	&	0.40	$\pm$	0.20	\\
060115	&	3.53	&	1.30E-07	$\pm$	1.60E-08	&	$\cdots$	&	$\cdots$	&	$\cdots$	&	$	62	\pm_{	19	}^{	 6	}$	&	0.40	$\pm$	0.20	\\
060116	&	6.60	&	2.00E-07	$\pm$	1.10E-07	&	$\cdots$	&	$\cdots$	&	$\cdots$	&	$	139	\pm_{	400	}^{	 36	}$	&	1.30	$\pm$	0.50	\\
060124	&	2.30	&	1.10E-06	$\pm$	1.20E-07	&	3.37E-05	$\pm$	3.40E-06	&	0.0021	$\pm$	0.0002	&	0.08	 $\pm$	0.04	&	$	237	\pm_{	76	}^{	51	}$	&	0.30	$\pm$	0.10	 \\
060206	&	4.05	&	4.40E-07	$\pm$	1.90E-08	&	$\cdots$	&	$\cdots$	&	0.10	$\pm$	0.10	&	$	75	 \pm_{	12	}^{	12	}$	&	1.25	$\pm$	0.25	\\
060210	&	3.91	&	5.50E-07	$\pm$	2.20E-08	&	1.94E-05	$\pm$	1.20E-06	&	0.0005	$\pm$	0.0001	&	0.13	 $\pm$	0.08	&	$	149	\pm_{	400	}^{	35	}$	&	0.50	$\pm$	0.20	 \\
060223	&	4.41	&	2.10E-07	$\pm$	3.70E-08	&	$\cdots$	&	$\cdots$	&	0.38	$\pm$	0.10	&	$	71	 \pm_{	100	}^{	10	}$	&	0.50	$\pm$	0.10	\\
060418	&	1.49	&	1.50E-06	$\pm$	5.90E-08	&	$\cdots$	&	$\cdots$	&	0.26	$\pm$	0.06	&	$	230	 \pm_{	20	}^{	20	}$	&	0.32	$\pm$	0.08	\\
060502	&	1.51	&	3.70E-07	$\pm$	1.60E-07	&	$\cdots$	&	$\cdots$	&	3.50	$\pm$	0.50	&	$	156	 \pm_{	400	}^{	33	}$	&	3.10	$\pm$	0.30	\\
060510	&	4.90	&	1.00E-07	$\pm$	1.70E-08	&	$\cdots$	&	$\cdots$	&	$\cdots$	&	$	95	\pm_{	60	}^{	 30	}$	&	$\cdots$	\\
060526	&	3.21	&	2.40E-07	$\pm$	3.30E-08	&	1.17E-06	$\pm$	1.70E-07	&	0.0034	$\pm$	0.0014	&	0.13	 $\pm$	0.03	&	$	25	\pm_{	5	}^{	5	}$	&	0.20	$\pm$	0.05	 \\
060604	&	2.68	&	9.00E-08	$\pm$	1.60E-08	&	$\cdots$	&	$\cdots$	&	5.00	$\pm$	1.00	&	$	40	 \pm_{	5	}^{	5	}$	&	0.60	$\pm$	0.20	\\
060605	&	3.80	&	1.20E-07	$\pm$	5.50E-08	&	$\cdots$	&	$\cdots$	&	5.00	$\pm$	3.00	&	$	169	 \pm_{	200	}^{	30	}$	&	2.00	$\pm$	0.50	\\
060607	&	3.08	&	2.70E-07	$\pm$	8.10E-08	&	$\cdots$	&	$\cdots$	&	2.00	$\pm$	0.50	&	$	120	 \pm_{	190	}^{	17	}$	&	2.00	$\pm$	0.20	\\
060707	&	3.43	&	1.53E-07	$\pm$	2.12E-08	&	3.41E-06	$\pm$	1.96E-07	&	$\cdots$	&	$\cdots$	&	$	 63	\pm_{	13	}^{	6	}$	&	$\cdots$	\\
060714	&	2.71	&	2.30E-07	$\pm$	1.42E-08	&	6.88E-06	$\pm$	2.47E-07	&	$\cdots$	&	$\cdots$	&	$	 103	\pm_{	21	}^{	16	}$	&	$\cdots$	\\
060729	&	0.54	&	1.93E-07	$\pm$	1.30E-08	&	6.43E-06	$\pm$	3.16E-07	&	$\cdots$	&	$\cdots$	&	$	 61	\pm_{	9	}^{	9	}$	&	$\cdots$	\\
060814	&	0.84	&	1.83E-06	$\pm$	4.44E-08	&	4.94E-05	$\pm$	4.91E-07	&	$\cdots$	&	0.29	$\pm$	 0.03	&	$	257	\pm_{	74	}^{	35	}$	&	1.65	$\pm$	0.24	\\
060904B	&	0.70	&	4.37E-07	$\pm$	2.28E-08	&	4.05E-06	$\pm$	2.17E-07	&	$\cdots$	&	0.36	$\pm$	 0.09	&	$	80	\pm_{	770	}^{	12	}$	&	1.00	$\pm$	0.16	\\
060908	&	2.43	&	6.69E-07	$\pm$	3.36E-08	&	7.68E-06	$\pm$	1.85E-07	&	$\cdots$	&	0.26	$\pm$	 0.06	&	$	151	\pm_{	112	}^{	25	}$	&	0.52	$\pm$	0.09	\\
060926	&	3.21	&	1.56E-07	$\pm$	1.22E-08	&	5.47E-07	$\pm$	3.80E-08	&	$\cdots$	&	1.03	$\pm$	 0.11	&	$	20	\pm_{	11	}^{	11	}$	&	$\cdots$	\\
060927	&	5.60	&	4.02E-07	$\pm$	1.54E-08	&	2.37E-06	$\pm$	8.67E-08	&	$\cdots$	&	0.12	$\pm$	 0.04	&	$	72	\pm_{	15	}^{	7	}$	&	0.46	$\pm$	0.12	\\
061007	&	1.26	&	7.20E-06	$\pm$	1.11E-07	&	2.24E-04	$\pm$	1.72E-06	&	$\cdots$	&	0.11	$\pm$	 0.01	&	$	399	\pm_{	12	}^{	11	}$	&	0.38	$\pm$	0.02	\\
061110A	&	0.76	&	9.79E-08	$\pm$	1.35E-08	&	2.71E-06	$\pm$	1.18E-07	&	$\cdots$	&	$\cdots$	&	$	 90	\pm_{	13	}^{	13	}$	&	$\cdots$	\\
061110B	&	3.44	&	1.79E-07	$\pm$	2.66E-08	&	6.12E-06	$\pm$	3.38E-07	&	$\cdots$	&	0.24	$\pm$	 0.36	&	$	517	\pm_{	53	}^{	53	}$	&	0.79	$\pm$	0.64	\\
061121	&	1.31	&	8.04E-06	$\pm$	1.07E-07	&	6.53E-05	$\pm$	5.76E-07	&	$\cdots$	&	0.03	$\pm$	 0.01	&	$	606	\pm_{	55	}^{	44	}$	&	0.98	$\pm$	0.19	\\
061222B	&	3.36	&	2.29E-07	$\pm$	3.15E-08	&	5.01E-06	$\pm$	2.49E-07	&	$\cdots$	&	$\cdots$	&	$	 49	\pm_{	8	}^{	8	}$	&	$\cdots$	\\
070110	&	2.35	&	1.12E-07	$\pm$	1.36E-08	&	4.04E-06	$\pm$	1.64E-07	&	$\cdots$	&	$\cdots$	&	$	 110	\pm_{	30	}^{	30	}$	&	$\cdots$	\\
070208	&	1.17	&	1.39E-07	$\pm$	2.06E-08	&	1.06E-06	$\pm$	1.46E-07	&	$\cdots$	&	$\cdots$	&	$	 51	\pm_{	10	}^{	10	}$	&	$\cdots$	\\
070318	&	0.84	&	4.10E-07	$\pm$	2.12E-08	&	7.34E-06	$\pm$	2.01E-07	&	$\cdots$	&	$\cdots$	&	$	 154	\pm_{	19	}^{	19	}$	&	0.72	$\pm$	0.24	\\
070411	&	2.95	&	1.50E-07	$\pm$	1.31E-08	&	6.29E-06	$\pm$	2.19E-07	&	$\cdots$	&	$\cdots$	&	$	 83	\pm_{	11	}^{	11	}$	&	$\cdots$	\\
070506	&	2.31	&	1.67E-07	$\pm$	1.38E-08	&	5.16E-07	$\pm$	3.43E-08	&	$\cdots$	&	2.52	$\pm$	 0.04	&	$	31	\pm_{	2	}^{	3	}$	&	0.12	$\pm$	0.06	\\
070508	&	0.82	&	7.67E-06	$\pm$	1.18E-07	&	7.26E-05	$\pm$	6.15E-07	&	$\cdots$	&	0.04	$\pm$	 0.01	&	$	233	\pm_{	7	}^{	7	}$	&	0.20	$\pm$	0.01	\\
070521	&	0.55	&	2.09E-06	$\pm$	5.26E-08	&	2.97E-05	$\pm$	4.00E-07	&	$\cdots$	&	0.04	$\pm$	 0.01	&	$	222	\pm_{	16	}^{	12	}$	&	0.58	$\pm$	0.06	\\
070529	&	2.50	&	3.32E-07	$\pm$	5.08E-08	&	7.44E-06	$\pm$	4.31E-07	&	$\cdots$	&	$\cdots$	&	$	 180	\pm_{	52	}^{	52	}$	&	$\cdots$	\\
070611	&	2.04	&	1.45E-07	$\pm$	2.25E-08	&	9.52E-07	$\pm$	8.44E-08	&	$\cdots$	&	$\cdots$	&	$	 92	\pm_{	30	}^{	30	}$	&	$\cdots$	\\
070612A	&	0.62	&	2.77E-07	$\pm$	4.24E-08	&	2.72E-05	$\pm$	9.37E-07	&	$\cdots$	&	$\cdots$	&	$	 87	\pm_{	17	}^{	17	}$	&	2.49	$\pm$	1.48	\\
070714B	&	0.92	&	3.24E-06	$\pm$	1.46E-07	&	8.91E-06	$\pm$	6.77E-07	&	$\cdots$	&	0.03	$\pm$	 0.01	&	$	1120	\pm_{	473	}^{	230	}$	&	0.45	$\pm$	0.04	\\
070802	&	2.45	&	6.38E-08	$\pm$	9.69E-09	&	6.50E-07	$\pm$	7.05E-08	&	$\cdots$	&	$\cdots$	&	$	 70	\pm_{	25	}^{	25	}$	&	$\cdots$	\\
070810A	&	2.17	&	2.77E-07	$\pm$	1.77E-08	&	1.59E-06	$\pm$	8.43E-08	&	$\cdots$	&	1.09	$\pm$	 0.23	&	$	44	\pm_{	9	}^{	9	}$	&	0.73	$\pm$	0.22	\\
071003	&	1.10	&	4.71E-06	$\pm$	1.82E-07	&	6.73E-05	$\pm$	1.48E-06	&	$\cdots$	&	0.38	$\pm$	 0.05	&	$	799	\pm_{	75	}^{	61	}$	&	0.88	$\pm$	0.07	\\
071010A	&	0.98	&	1.17E-07	$\pm$	2.67E-08	&	4.97E-07	$\pm$	6.05E-08	&	$\cdots$	&	$\cdots$	&	$	 27	\pm_{	10	}^{	10	}$	&	$\cdots$	\\
071010B	&	0.95	&	9.20E-07	$\pm$	2.18E-08	&	8.37E-06	$\pm$	1.16E-07	&	$\cdots$	&	0.84	$\pm$	 0.04	&	$	52	\pm_{	6	}^{	8	}$	&	1.21	$\pm$	0.03	\\
071031	&	2.69	&	7.08E-08	$\pm$	8.61E-09	&	2.19E-06	$\pm$	1.92E-07	&	$\cdots$	&	$\cdots$	&	$	 24	\pm_{	7	}^{	7	}$	&	$\cdots$	\\
071117	&	1.33	&	2.71E-06	$\pm$	5.83E-08	&	7.97E-06	$\pm$	2.02E-07	&	$\cdots$	&	0.60	$\pm$	 0.01	&	$	278	\pm_{	143	}^{	48	}$	&	0.20	$\pm$	0.02	\\
071122	&	1.14	&	6.76E-08	$\pm$	2.06E-08	&	1.41E-06	$\pm$	1.63E-07	&	$\cdots$	&	$\cdots$	&	$	 73	\pm_{	30	}^{	30	}$	&	$\cdots$	\\
080210	&	2.64	&	2.57E-07	$\pm$	1.95E-08	&	4.17E-06	$\pm$	1.41E-07	&	$\cdots$	&	0.53	$\pm$	 0.17	&	$	73	\pm_{	15	}^{	15	}$	&	0.57	$\pm$	0.44	\\
080310	&	2.43	&	1.83E-07	$\pm$	1.72E-08	&	5.49E-06	$\pm$	2.90E-07	&	$\cdots$	&	$\cdots$	&	$	 28	\pm_{	6	}^{	6	}$	&	0.41	$\pm$	0.55	\\
080319B	&	0.94	&	1.55E-05	$\pm$	1.91E-07	&	5.25E-04	$\pm$	3.94E-06	&	$\cdots$	&	0.02	$\pm$	 0.01	&	$	651	\pm_{	8	}^{	8	}$	&	0.14	$\pm$	0.01	\\
080319C	&	1.95	&	2.22E-06	$\pm$	7.79E-08	&	1.77E-05	$\pm$	2.99E-07	&	$\cdots$	&	$\cdots$	&	$	 307	\pm_{	85	}^{	56	}$	&	0.21	$\pm$	0.12	\\
080330	&	1.51	&	1.33E-07	$\pm$	1.80E-08	&	8.77E-07	$\pm$	1.26E-07	&	$\cdots$	&	$\cdots$	&	$	 20	\pm_{	9	}^{	9	}$	&	$\cdots$	\\
080411	&	1.03	&	1.04E-05	$\pm$	1.31E-07	&	8.75E-05	$\pm$	2.01E-07	&	$\cdots$	&	0.21	$\pm$	 0.01	&	$	259	\pm_{	21	}^{	16	}$	&	0.65	$\pm$	0.01	\\
080413A	&	2.43	&	1.22E-06	$\pm$	2.65E-08	&	9.86E-06	$\pm$	1.71E-07	&	$\cdots$	&	0.13	$\pm$	 0.03	&	$	170	\pm_{	48	}^{	24	}$	&	0.23	$\pm$	0.03	\\
080413B	&	1.10	&	3.17E-06	$\pm$	8.25E-08	&	8.00E-06	$\pm$	1.52E-07	&	$\cdots$	&	0.23	$\pm$	 0.01	&	$	73	\pm_{	10	}^{	10	}$	&	0.50	$\pm$	0.03	\\
080430	&	0.77	&	4.60E-07	$\pm$	2.15E-08	&	3.01E-06	$\pm$	1.53E-07	&	$\cdots$	&	0.68	$\pm$	 0.08	&	$	80	\pm_{	15	}^{	15	}$	&	0.76	$\pm$	0.12	\\
080516	&	3.20	&	2.77E-07	$\pm$	2.80E-08	&	5.88E-07	$\pm$	5.50E-08	&	$\cdots$	&	0.15	$\pm$	 0.01	&	$	66	\pm_{	24	}^{	24	}$	&	$\cdots$	\\
080520	&	1.55	&	8.23E-08	$\pm$	1.00E-08	&	1.59E-07	$\pm$	3.00E-08	&	$\cdots$	&	$\cdots$	&	$	 12	\pm_{	5	}^{	5	}$	&	$\cdots$	\\
080603B	&	2.69	&	7.57E-07	$\pm$	2.63E-08	&	7.02E-06	$\pm$	1.78E-07	&	$\cdots$	&	0.08	$\pm$	 0.01	&	$	85	\pm_{	55	}^{	18	}$	&	0.22	$\pm$	0.03	\\
080605	&	1.64	&	5.99E-06	$\pm$	1.10E-07	&	4.72E-05	$\pm$	4.32E-07	&	$\cdots$	&	0.11	$\pm$	 0.01	&	$	246	\pm_{	14	}^{	11	}$	&	0.22	$\pm$	0.01	\\
080607	&	3.04	&	8.35E-06	$\pm$	2.42E-07	&	1.00E-04	$\pm$	0.00E+00	&	$\cdots$	&	0.04	$\pm$	 0.01	&	$	394	\pm_{	35	}^{	33	}$	&	0.18	$\pm$	0.06	\\
080707	&	1.23	&	1.68E-07	$\pm$	1.02E-08	&	1.26E-06	$\pm$	8.87E-08	&	$\cdots$	&	$\cdots$	&	$	 73	\pm_{	20	}^{	20	}$	&	$\cdots$	\\
080721	&	2.60	&	9.57E-06	$\pm$	5.01E-07	&	5.99E-05	$\pm$	3.04E-06	&	$\cdots$	&	0.13	$\pm$	 0.05	&	$	485	\pm_{	41	}^{	36	}$	&	0.09	$\pm$	0.04	\\
080810	&	3.35	&	9.76E-07	$\pm$	8.40E-08	&	1.80E-05	$\pm$	1.30E-06	&	$\cdots$	&	$\cdots$	&	$	 313.5	\pm_{	73.6	}^{	73.6	}$	&	$\cdots$	\\
080913	&	6.70	&	2.31E-07	$\pm$	4.00E-08	&	1.15E-06	$\pm$	1.20E-07	&	$\cdots$	&	$\cdots$	&	$	 93.1	\pm_{	56.1	}^{	56.1	}$	&	$\cdots$	\\
080916A	&	0.69	&	1.06E-06	$\pm$	1.60E-07	&	2.19E-05	$\pm$	7.30E-06	&	$\cdots$	&	$\cdots$	&	$	 109	\pm_{	9	}^{	9	}$	&	$\cdots$	\\
081121	&	2.51	&	1.87E-06	$\pm$	4.80E-07	&	1.73E-05	$\pm$	3.20E-06	&	$\cdots$	&	$\cdots$	&	$	 248	\pm_{	38	}^{	32	}$	&	$\cdots$	\\
081222	&	2.77	&	2.62E-06	$\pm$	3.80E-07	&	1.70E-05	$\pm$	1.40E-06	&	$\cdots$	&	$\cdots$	&	$	 134	\pm_{	9	}^{	9	}$	&	$\cdots$	\\
090102	&	1.55	&	3.90E-06	$\pm$	5.60E-07	&	3.66E-05	$\pm$	3.40E-06	&	$\cdots$	&	$\cdots$	&	$	 451	\pm_{	73	}^{	58	}$	&	$\cdots$	\\
090323	&	3.57	&	3.80E-06	$\pm$	6.90E-07	&	1.48E-04	$\pm$	2.00E-05	&	$\cdots$	&	$\cdots$	&	$	 416	\pm_{	76	}^{	73	}$	&	$\cdots$	\\
090328	&	0.74	&	7.28E-06	$\pm$	5.20E-07	&	1.37E-04	$\pm$	8.00E-06	&	$\cdots$	&	$\cdots$	&	$	 653	\pm_{	45	}^{	45	}$	&	$\cdots$	\\
090423	&	8.20	&	3.24E-07	$\pm$	6.80E-08	&	1.17E-06	$\pm$	3.20E-07	&	$\cdots$	&	$\cdots$	&	$	 82	\pm_{	15	}^{	15	}$	&	$\cdots$	\\
090424	&	0.54	&	1.80E-05	$\pm$	1.20E-06	&	5.85E-05	$\pm$	2.10E-06	&	$\cdots$	&	$\cdots$	&	$	 177	\pm_{	3	}^{	3	}$	&	$\cdots$	\\
090516	&	4.11	&	5.93E-07	$\pm$	3.40E-08	&	1.09E-05	$\pm$	1.60E-06	&	$\cdots$	&	$\cdots$	&	$	 185.6	\pm_{	98.4	}^{	42.5	}$	&	$\cdots$	\\
090618	&	0.54	&	8.58E-06	$\pm$	9.60E-07	&	3.39E-04	$\pm$	2.50E-05	&	$\cdots$	&	$\cdots$	&	$	 155.5	\pm_{	11.1	}^{	10.5	}$	&	$\cdots$	\\
090715B	&	3.00	&	8.96E-07	$\pm$	2.49E-07	&	1.09E-05	$\pm$	1.60E-06	&	$\cdots$	&	$\cdots$	&	$	 134	\pm_{	56	}^{	30	}$	&	$\cdots$	\\
090812	&	2.45	&	2.17E-06	$\pm$	2.40E-07	&	3.15E-05	$\pm$	4.10E-06	&	$\cdots$	&	$\cdots$	&	$	 572	\pm_{	251	}^{	159	}$	&	$\cdots$	\\
090902B	&	1.82	&	2.72E-05	$\pm$	3.00E-07	&	3.78E-04	$\pm$	3.00E-06	&	$\cdots$	&	$\cdots$	&	$	 775	\pm_{	11	}^{	11	}$	&	$\cdots$	\\
090926	&	2.11	&	1.82E-05	$\pm$	3.00E-07	&	1.51E-04	$\pm$	7.70E-06	&	$\cdots$	&	$\cdots$	&	$	 314	\pm_{	4	}^{	4	}$	&	$\cdots$	\\
090926B	&	1.24	&	5.55E-07	$\pm$	7.90E-08	&	1.66E-05	$\pm$	5.00E-07	&	$\cdots$	&	$\cdots$	&	$	 78.3	\pm_{	7	}^{	7	}$	&	$\cdots$	\\
091018	&	0.97	&	1.82E-06	$\pm$	8.40E-07	&	4.33E-06	$\pm$	7.60E-07	&	$\cdots$	&	$\cdots$	&	$	 19.2	\pm_{	18	}^{	11	}$	&	$\cdots$	\\
091020	&	1.71	&	1.63E-06	$\pm$	2.00E-07	&	1.68E-05	$\pm$	3.70E-06	&	$\cdots$	&	$\cdots$	&	$	 47.9	\pm_{	7.1	}^{	7.1	}$	&	$\cdots$	\\
091029	&	2.75	&	2.82E-07	$\pm$	1.60E-08	&	5.84E-06	$\pm$	4.70E-07	&	$\cdots$	&	$\cdots$	&	$	 61.4	\pm_{	17.5	}^{	17.5	}$	&	$\cdots$	\\
091127	&	0.49	&	4.03E-06	$\pm$	1.70E-07	&	2.65E-05	$\pm$	5.00E-07	&	$\cdots$	&	$\cdots$	&	$	 36	\pm_{	2	}^{	2	}$	&	$\cdots$	\\
091208B	&	1.06	&	3.47E-06	$\pm$	6.00E-07	&	7.78E-06	$\pm$	8.80E-07	&	$\cdots$	&	$\cdots$	&	$	 124	\pm_{	20.1	}^{	19.4	}$	&	$\cdots$	\\
\hline
\end{longtable}
\end{tiny}
\end{center}


\begin{figure}[pb]
\centerline{\psfig{file=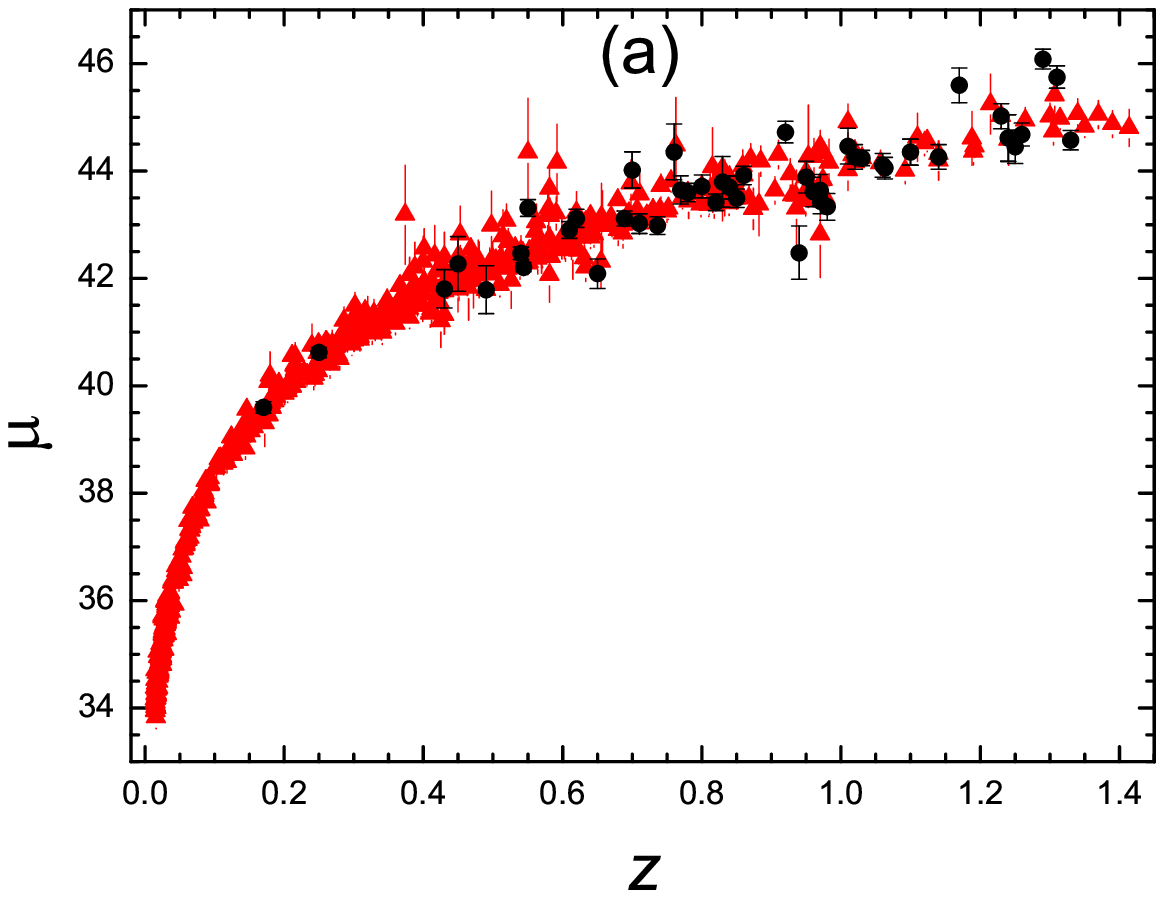,width=6.5cm,height=4.5cm}
            \psfig{file=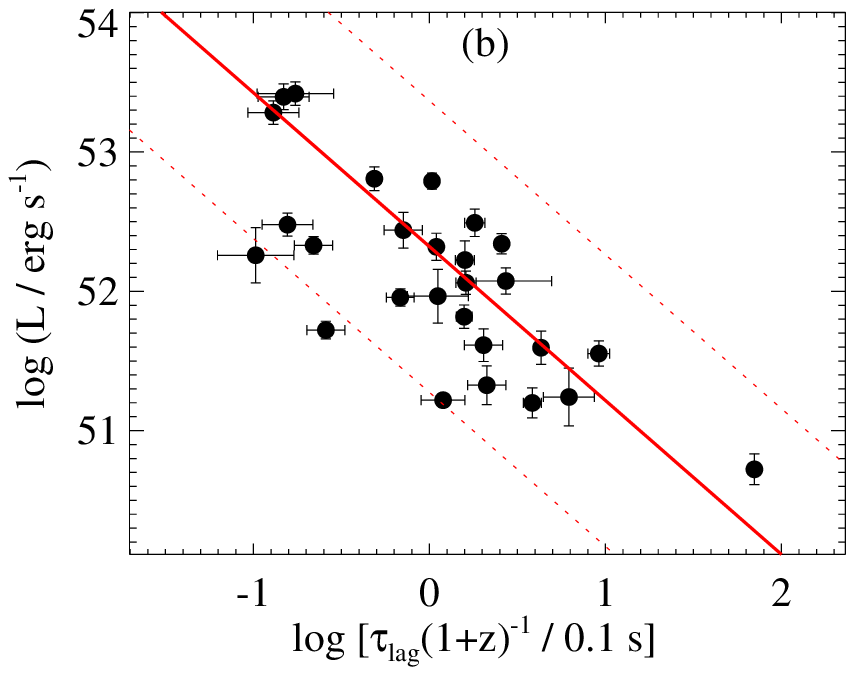,width=6cm}}
\centerline{\psfig{file=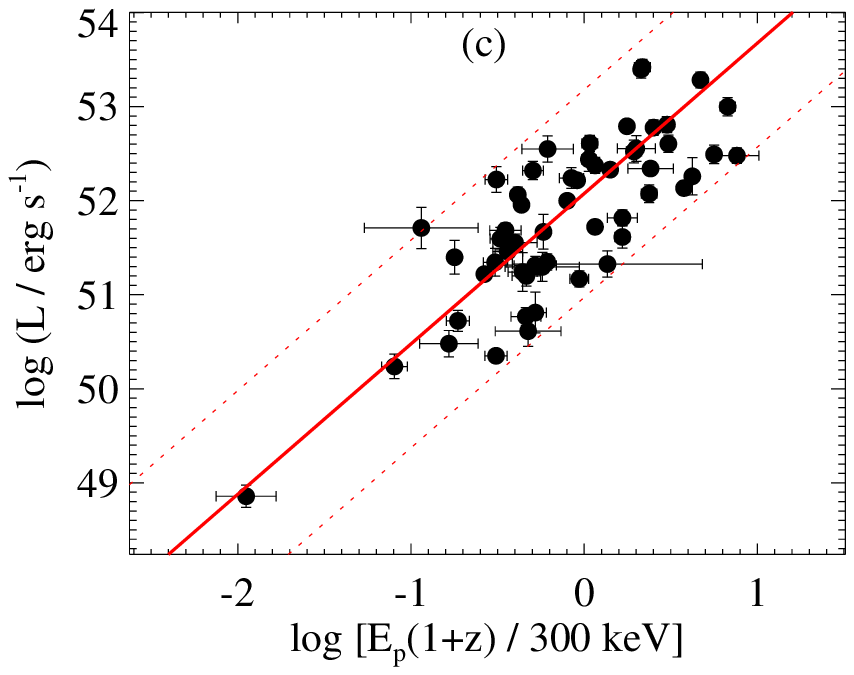,width=6cm}
            \psfig{file=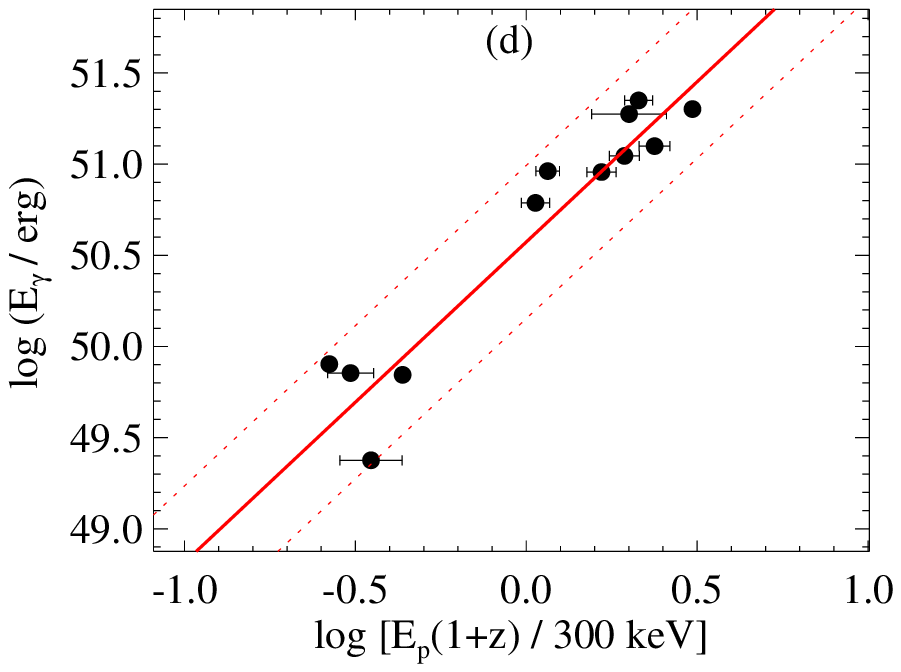,width=6cm}}
\centerline{\psfig{file=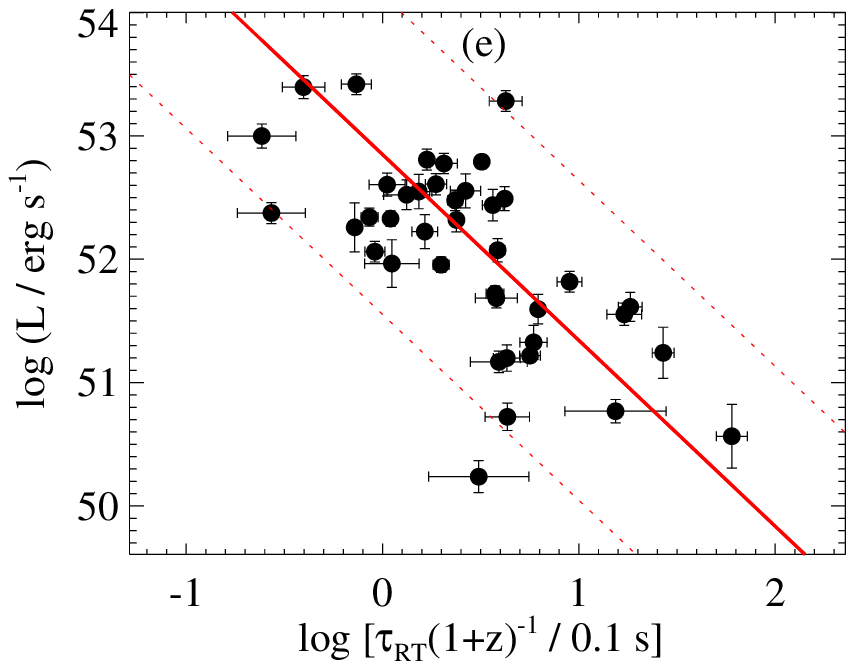,width=6cm}
            \psfig{file=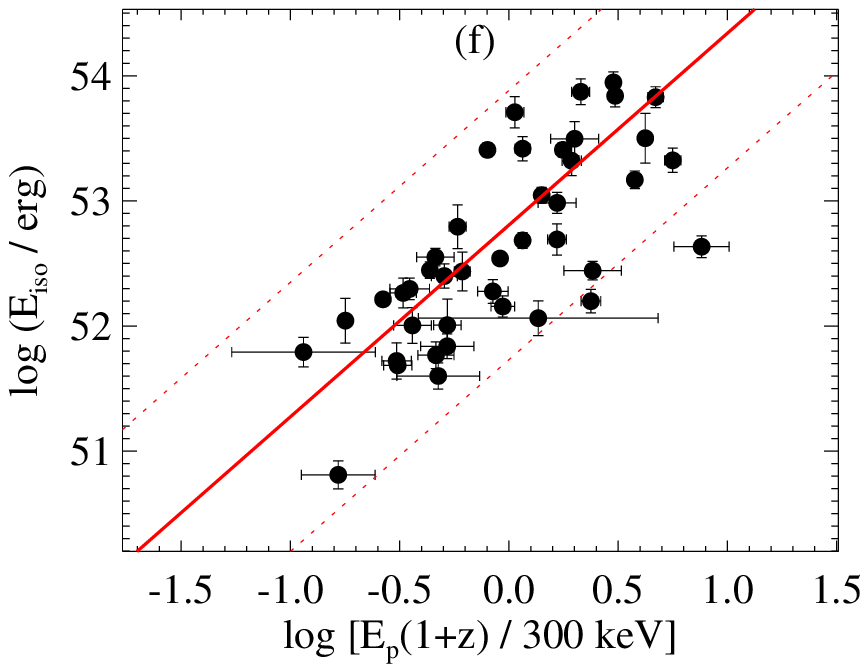,width=6cm}}
\vspace*{8pt}
\caption{(a): The Hubble diagram of 580 SNe Ia (red triangles) and 57 GRBs at $z\leq1.4$ (black dots)
whose distance moduli are derived by using cubic spline interpolation. (b)-(f): The $\tau_{\rm lag}-L$,
$E_{\rm p}-L$, $E_{\rm p}-E_{\gamma}$, $\tau_{\rm RT}-L$, and $E_{\rm p}-E_{\rm iso}$ correlations.
The five correlations are calibrated with the sample at $z\leq1.4$ using cubic spline interpolation.
The solid lines show the best-fitting results, while the dashed lines represent their 2 $\sigma$
dispersion around the best fits.\label{relation}}
\end{figure}

\subsection{Calibration}
First of all, we obtain the distance moduli of 57 low-\emph{z} ($z<1.4$) GRBs
by using cubic spline interpolation from the 580 Union2.1 SNe Ia compiled in Ref.~\cite{Suzuki12}.
The interpolated distance moduli $\mu$ of these 57 GRBs and their corresponding
errors $\sigma_{\mu}$ are shown in Fig.~\ref{relation}(a). The SNe Ia data are also
plotted in Fig.~\ref{relation}(a) for comparison. When the cubic spline interpolation
is used, the error of the interpolated distance modulus $\mu$ for the GRB at redshift \emph{z}
can be calculated by
\begin{equation}
\sigma_{\mu}=\left[\left(\frac{z_{i+1}-z}{z_{i+1}-z_{i}}\right)^{2}\epsilon_{\mu,i}^{2}
+\left(\frac{z-z_{i}}{z_{i+1}-z_{i}}\right)^{2}\epsilon_{\mu,i+1}^{2}\right]^{1/2},
\label{error}
\end{equation}
where $\epsilon_{\mu,i}$ and $\epsilon_{\mu,i+1}$ are errors of the SNe at nearby redshifts
$z_{i}$ and $z_{i+1}$, respectively. With $D_{\rm L}$ in units of Mpc, the
predicted distance modulus is defined as
\begin{equation}
\mu=5\log(D_{\rm L})+25.
\label{distance}
\end{equation}
Having estimated the distance moduli $\mu$ of 57 low-\emph{z} GRBs in a model independent way,
we can convert $\mu$ into luminosity distance $D_{\rm L}$ by using Eq.~(\ref{distance}).
From Eqs.~(\ref{relation})-(\ref{correctenergy}) with the corresponding $P_{\rm bolo}$,
$S_{\rm bolo}$, and $f_{\rm beam}$, we can calculate $L$, $E_{\rm iso}$, and $E_{\rm \gamma}$.
In Fig.~\ref{relation}(b)-\ref{relation}(f), with the interpolation results, we show the
five luminosity indicators for these 57 GRBs at $z<1.4$. For each relation, we perform a
linear least-squares fit, taking into account both the X axis error and the Y axis error.
We also measure the scatter of each relation with the distance of the data points from the
best-fit line, as done by Ref.~\cite{Ghirlanda05}. The best-fitting results of the intercept
$a$ and the slope $b$ with their $1\sigma$ uncertainties and the linear correlation coefficients
for each relation are summarized in Table~\ref{fitresult}. The best-fitting results derived by
using the interpolation method are carried out with these 57 GRBs at $z<1.4$. In other word,
the results are derived by using data from 27, 55, 12, 40, and 42 GRBs for the $\tau_{\rm lag}-L$,
$E_{\rm p}-L$, $E_{\rm p}-E_{\gamma}$, $\tau_{\rm RT}-L$, and $E_{\rm p}-E_{\rm iso}$ relations,
respectively.

\begin{table}[htb]
  \begin{center}
  \caption{Best-fitting results}
  \label{fitresult}
  \begin{tabular}{ccccc} \hline
Relation &   a   &   b   &    N    &     $r$   \\
\hline
$\tau_{\rm lag}-L$      &	52.32 $\pm$ 0.03	&	-1.10 $\pm$ 0.05	&	27	&	-0.75	\\
$E_{\rm p}-L$           &	52.08 $\pm$ 0.02	&	1.60  $\pm$ 0.04	&	55	&	0.79	\\
$E_{\rm p}-E_{\gamma}$  &	50.57 $\pm$ 0.01	&	1.76  $\pm$ 0.03	&	12	&	0.95	\\
$\tau_{\rm RT}-L$       &	52.85 $\pm$ 0.03	&	-1.51 $\pm$ 0.06	&	40	&	-0.66	\\
$E_{\rm p}-E_{\rm iso}$ &	52.81 $\pm$ 0.02	&	1.53  $\pm$ 0.04	&	42	&	0.73	\\
\hline
\end{tabular}
\end{center}
\end{table}

Ref.~\cite{Wang11} found no statistically significant evidence for the redshift evolution
of the luminosity relations. If the GRB luminosity relations indeed do not evolve with redshift,
we can extend the calibrated luminosity relations to high-\emph{z} ($z>1.4$) and derive the
luminosity ($L$) or energy ($E_{\rm iso}$ or $E_{\rm \gamma}$) of each burst at high-\emph{z}
by utilizing the calibrated relations. Therefore, the luminosity distance $D_{\rm L}$
can be derived from Eqs.~(\ref{relation})-(\ref{correctenergy}). The uncertainty of
the value of the luminosity or energy deduced from each relation is
\begin{equation}
\sigma_{\log y}^{2}=\sigma_{a}^{2}+\left(\sigma_{b}^{2}\log x\right)^{2}+
\left(\frac{b}{\ln 10}\frac{\sigma_{x}}{x}\right)^{2}+\sigma_{\rm int}^{2},
\end{equation}
where $\sigma_{a}$, $\sigma_{b}$, and $\sigma_{x}$ are $1\sigma$ uncertainties of the intercept
$a$, the slope $b$, and the GRB measurable parameters $x$, and $\sigma_{\rm int}$ is the
systematic error in the fitting that accounts for the extra scatter of the luminosity relations.
Then, we obtain the distance moduli $\mu$ for these 82 GRBs at $z>1.4$ using Eq.~(\ref{distance}).
The propagated uncertainties will depend on whether $P_{\rm bolo}$ or $S_{\rm bolo}$ are given by
\begin{equation}
\sigma_{\mu}=\left[\left(\frac{5}{2}\sigma_{\log L}\right)^{2}
+\left(\frac{5}{2\ln 10}\frac{\sigma_{P_{\rm bolo}}}{P_{\rm bolo}}\right)^{2}\right]^{1/2},
\end{equation}
or
\begin{equation}
\sigma_{\mu}=\left[\left(\frac{5}{2}\sigma_{\log E_{\rm iso}}\right)^{2}
+\left(\frac{5}{2\ln 10}\frac{\sigma_{S_{\rm bolo}}}{S_{\rm bolo}}\right)^{2}\right]^{1/2},
\end{equation}
and
\begin{equation}
\sigma_{\mu}=\left[\left(\frac{5}{2}\sigma_{\log E_{\gamma}}\right)^{2}
+\left(\frac{5}{2\ln 10}\frac{\sigma_{S_{\rm bolo}}}{S_{\rm bolo}}\right)^{2}
+\left(\frac{5}{2\ln 10}\frac{\sigma_{f_{\rm beam}}}{f_{\rm beam}}\right)^{2}\right]^{1/2}.
\end{equation}
Here we ignore the uncertainty of \emph{z} in our calculations.

After obtaining the distance modulus of each GRB using one of these relations,
we use the same method as Ref.~\cite{Schaefer07} to calculate the real distance
modulus, which is the weighted average of all available distance modulus.
The real distance modulus for each burst is
\begin{equation}
\mu_{\rm fit}=\frac{\sum_{i}\mu_{i}/\sigma_{\mu_{i}}^{2}}{\sum_{i}\sigma_{\mu_{i}}^{-2}},
\end{equation}
with its corresponding uncertainty $\sigma_{\mu_{\rm fit}}=(\sum_{i}\sigma_{\mu_{i}}^{-2})^{-1/2}$,
where the summation runs from 1 to 5 over the relations with available data,
$\mu_{i}$ and $\sigma_{\mu_{i}}$ are the best estimated distance modulus
and its corresponding uncertainty from the $i$th relation.

\begin{figure}[htb]
\begin{center}
\vspace*{-1.0cm} \resizebox{10.0cm}{!}{\includegraphics{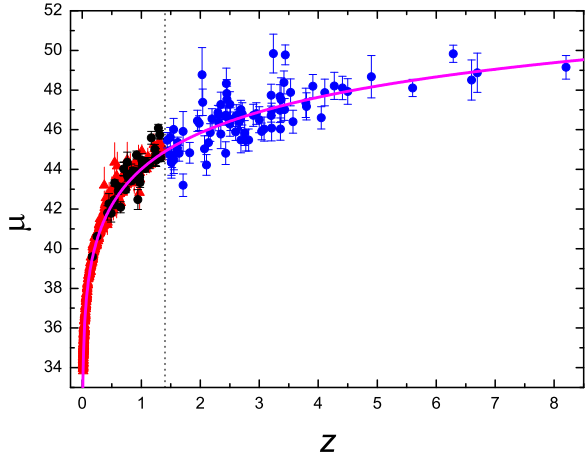}}
\caption{Hubble diagram of 580 SNIa (red triangles) and 139 GRBs (dots) obtained using the
   interpolation method. The 57 GRBs at $z\leq1.4$ are obtained by interpolating from SNe Ia
   data (black dots); and the 82 GRBs at $z>1.4$ (blue dots) are obtained with the five
   correlations calibrated with the sample at $z\leq1.4$ using the cubic spline interpolation
   method. The vertical dotted line represents $z=1.4$. The solid curve represents the best-fit
   cosmology for a flat $\Lambda$CDM universe: $\Omega_{m}=0.29$, $\Omega_{\Lambda}=0.71$. \label{HD}}
\end{center}
\end{figure}

Fig.~\ref{HD} shows the Hubble diagram from the Union2.1 SNe Ia sample
and 139 GRBs. The combined Hubble diagram is consistent with the concordance
cosmology. The 57 GRBs at $z<1.4$ are obtained using interpolation
method directly from SNe data. The 82 GRBs at $z>1.4$ are obtained by
utilizing the five relations calibrated with the sample at $z<1.4$ using
the cubic spline interpolation method.

\section{Constraints from supernovae and GRBs}

The latest Type Ia SNe dataset known as the Union2.1 sample was recently released
by the Supernova Cosmology Project collaboration, which contains 580 SNe detections
(see \cite{Suzuki12}). With luminosity distance $D_{\rm L}(\xi,z)$ in units of Mpc (where $\xi$
stands for all the cosmological parameters that define the fitted model), the theoretical
distance modulus $\mu_{\rm th}$ can be calculated by using Eq.~(\ref{distance}).
The likelihood functions can be determined from the $\chi^{2}$ statistic,
\begin{equation}
\chi_{\rm SNe}^{2}=\sum_{i=1}^{N}\frac{[\mu_{\rm th}(z_{i})-\mu_{\rm obs}(z_{i})]^{2}}
{\sigma_{\rm lc}^{2}+\sigma_{\rm ext}^{2}+\sigma_{\rm sample}^{2}},
\end{equation}
where $\sigma_{\rm lc}$ is the propagated error from the covariance matrix of the
lightcurve fit, and $\mu_{\rm obs}$ is the observational distance modulus. The
uncertainties due to host galaxy peculiar velocities, Galactic extinction corrections,
and gravitational lensing are included in $\sigma_{\rm ext}$, and $\sigma_{\rm sample}$
is a floating dispersion term containing sample-dependent systematic errors. The confidence
regions can be found through marginalizing the likelihood functions over Hubble constant
$H_{0}$ (i.e., integrating the probability density $p\propto\exp(-\chi^{2}/2)$ for all
values of $H_{0}$).

Gamma-ray bursts (GRBs) are the most luminous transient events in the cosmos. Owing to
their high luminosity, GRBs can be detected out to the edge of the visible Universe,
constituting a powerful tool for constructing a Hubble diagram at high-\emph{z}. We use
the above calibration results obtained by using the interpolation methods directly from
SNe Ia data. The $\chi^{2}$ value for the 82 GRBs at $z>1.4$ is given by
\begin{equation}
\chi_{\rm GRB}^{2}=\sum_{i=1}^{N}\frac{[\mu_{\rm th}(z_{i})-\mu_{\rm fit,\emph{i}}]^{2}}
{\sigma_{\mu_{\rm fit,\emph{i}}}^{2}},
\end{equation}
where $\mu_{\rm fit,\emph{i}}$ and $\sigma_{\mu_{\rm fit,\emph{i}}}$ are the fitted distance modulus
and its error for each burst. We also marginalize the nuisance parameter $H_{0}$.

Motivated by these significant updates in the observations of SNe Ia and GRBs, it is
natural to consider the joint constraints on cosmological parameters and dark energy
with the latest observational data. We combine SNe Ia and GRBs by multiplying the
likelihood functions. The total $\chi^{2}$ value is
\begin{equation}
\chi_{\rm total}^{2}=\chi_{\rm SNe}^{2}+\chi_{\rm GRB}^{2}.
\end{equation}
The best-fitting values of cosmological model are obtained by minimizing $\chi_{\rm total}^{2}$.

\subsection{$\Lambda$CDM model}
In a Friedmann-Robertson-Walker (FRW) cosmology with mass density $\Omega_{m}$ and
vacuum energy density $\Omega_{\Lambda}$, the luminosity distance is given as
\begin{equation}
D_{L}(z)={c\over H_{0}}{(1+z)\over\sqrt{\mid\Omega_{k}\mid}}\; \rm sinn\left\{\mid\Omega_{\emph{k}}\mid^{1/2}
\times\int_{0}^{\emph{z}}{d\emph{z}\over\sqrt{\Omega_{\emph{m}}(1+\emph{z})^{3}+\Omega_{\Lambda}+\Omega_{\emph{k}}(1+\emph{z})^{2}}}\right\}\;,
\label{FRW}
\end{equation}
where $c$ is the speed of light, $H_{0}$ is the Hubble constant at the present time,
$\Omega_{k}=1-\Omega_{m}-\Omega_{\Lambda}$ represents the spatial curvature of the
Universe, and sinn is $\sinh$ when $\Omega_{k}>0$ and $\sin$ when $\Omega_{k}<0$.
For a flat Universe with $\Omega_{k}=0$, Eq.~(\ref{FRW}) simplifies to the
form $(1+z)c/H_{0}$ times the integral. In this $\Lambda$CDM model, the transition
redshift satisfies
\begin{equation}
z_{T}=\left(\frac{2\Omega_{\Lambda}}{\Omega_{m}}\right)^{1/3}-1.
\end{equation}

\begin{figure}[htb]
\begin{center}
\begin{tabular}{cc}
\resizebox{8.7cm}{!}{\includegraphics{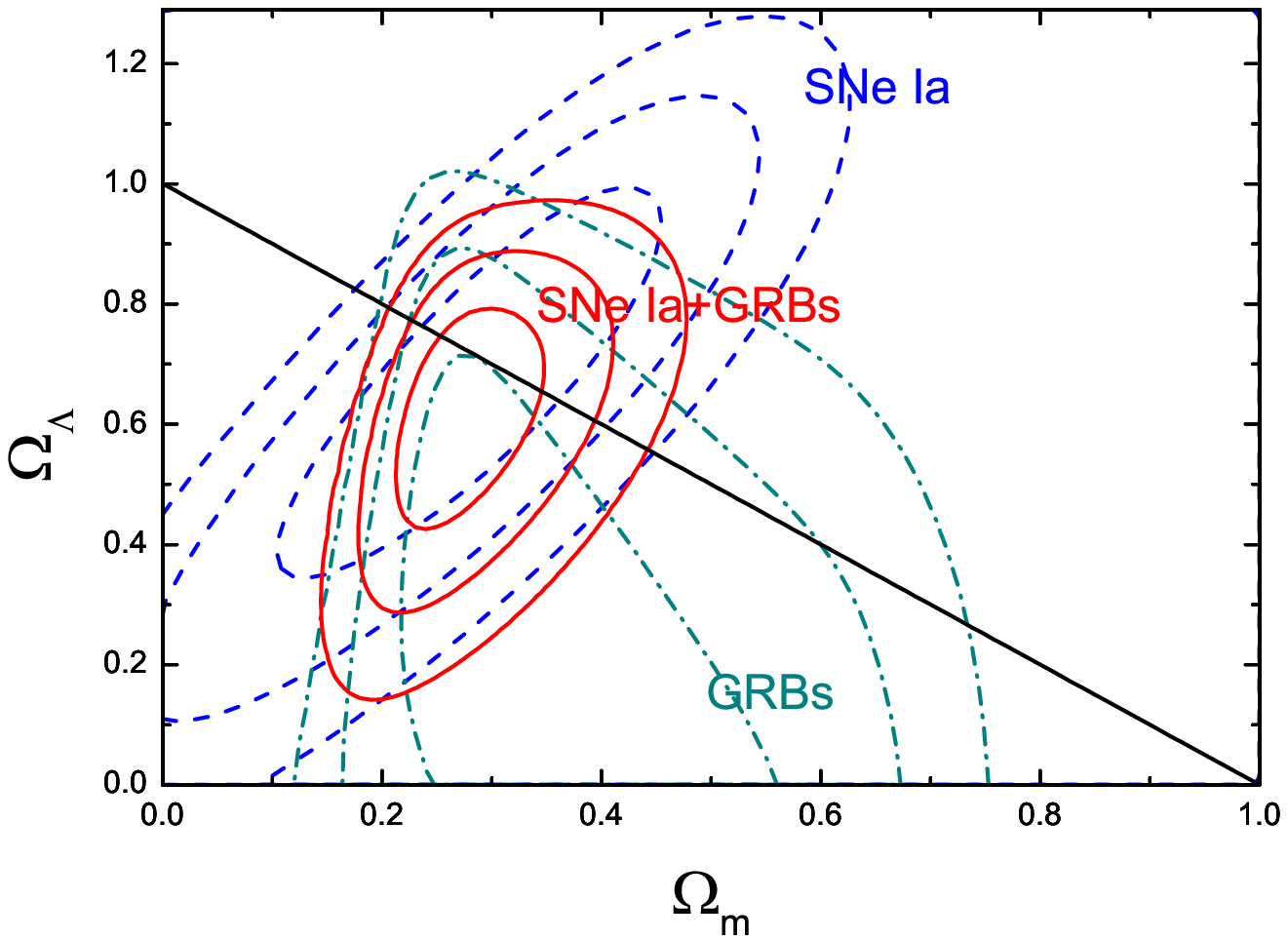}}&
\hspace*{-1.0cm} \resizebox{8.7cm}{!}{\includegraphics{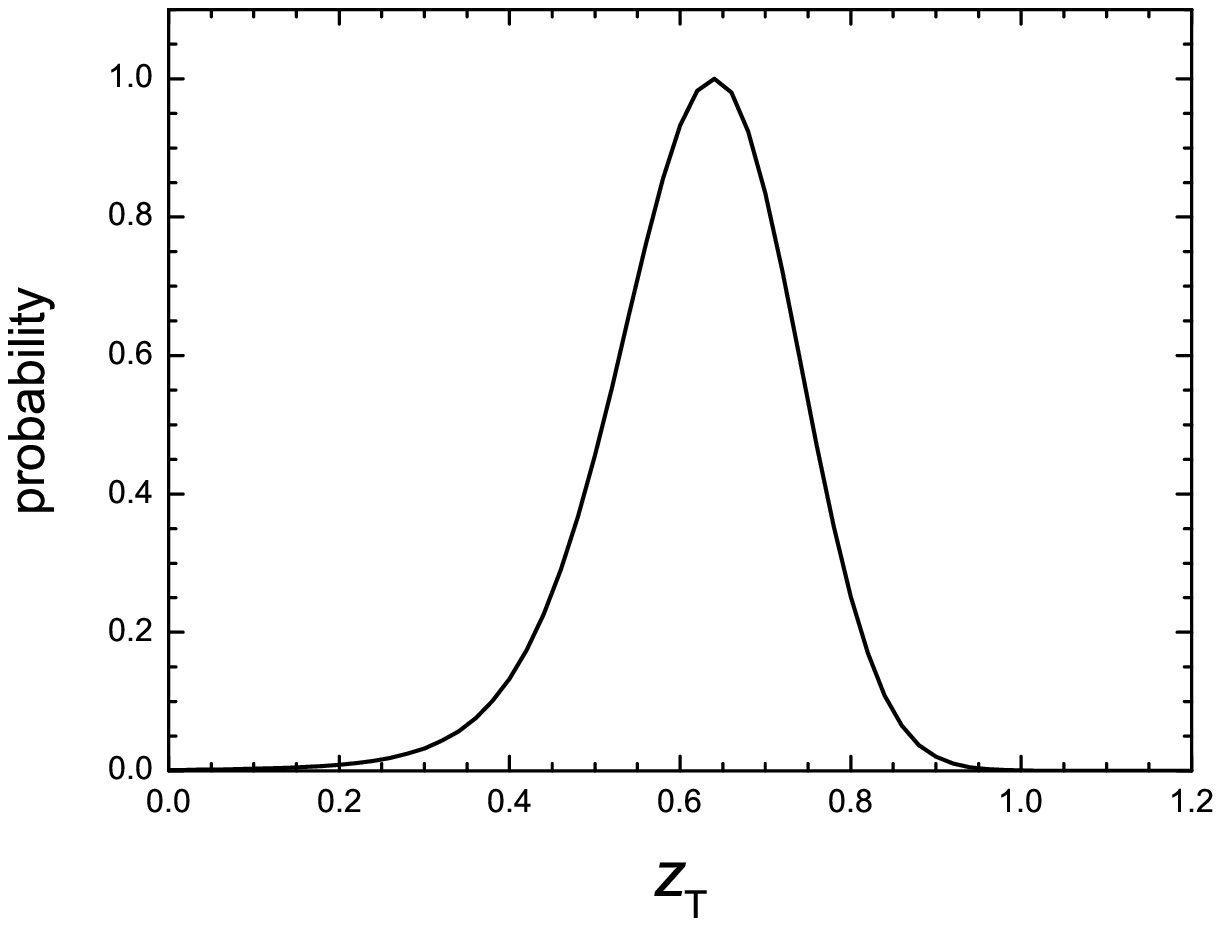}}\\
\end{tabular}
\caption{Left: The contour confidence levels of $(\Omega_{m}, \Omega_{\Lambda})$ in the $\Lambda$CDM
   model from the data for 82 GRBs ($z>1.4$) (dark cyan dash-dotted lines), 580 SNe Ia (blue dashed lines),
   and 82 GRBs $+$ 580 SNe Ia (red solid lines), respectively. The contours correspond to 1, 2, and $3\sigma$
   confidence regions. Right: The probability versus the transition redshift derived from the GRB
   and SNe Ia sample.
\label{LCDM}}
\end{center}
\end{figure}

We use the data sets discussed above to constrain cosmological parameters. In the
left panel of Fig.~\ref{LCDM}, we show the confidence regions for $(\Omega_{m},\Omega_{\Lambda})$
from 82 GRBs (dark cyan dash-dotted lines), 580 SNe Ia (blue dotted lines), and
82 GRBs $+$ 580 SNe Ia (red solid lines), respectively. We can see that adding 82 high
redshift GRBs ($z>1.4$) to 580 SNe Ia ($z<1.4$) significantly improves the constraint
on $\Omega_{m}-\Omega_{\Lambda}$ plane. The 1 $\sigma$ confidence region from all the
data sets is $(\Omega_{m},\Omega_{\Lambda})=(0.27_{-0.06}^{+0.08}, 0.62_{-0.19}^{+0.18})$
with $\chi_{\rm min}^{2}=727.01$ for 659 degrees of freedom. Under the assumption of
a flat universe (solid line), the contours yield $(\Omega_{m}, \Omega_{\Lambda})
=(0.29_{-0.04}^{+0.04}, 0.71_{-0.04}^{+0.04})$. The transition redshift at which the
Universe switched from deceleration to acceleration phase is $z_{T}=0.64_{-0.14}^{+0.08}$
at the $1\sigma$ confidence level (the right panel of Fig.~\ref{LCDM}).

\subsection{Cardassian Expansion Model}
Ref.~\cite{Freese02} proposed the Cardassian expansion model as a possible alternative
for explaining the acceleration of the Universe that invokes no vacuum energy. This model
allows an acceleration in a flat, matter-dominated cosmology. If we consider a spatially
flat FRW Universe, the Friedmann equation is modified as
\begin{equation}
H^{2}=\frac{8\pi G}{3}(\rho+C\rho^{n}).
\end{equation}
This modification may arise as a consequence of embedding our observable Universe
as a (3+1) dimensional brane in extra dimensions or the self-interaction of dark matter.
The luminosity distance in this model is given by
\begin{equation}
D_{L}(z)=cH_{0}^{-1}(1+z)\int_{0}^{z}dz[(1+z)^{3}\Omega_{m}+(1-\Omega_{m})(1+z)^{3n}]^{-1/2}.
\end{equation}

\begin{figure}[htb]
\begin{center}
 \resizebox{10.0cm}{!}{\includegraphics{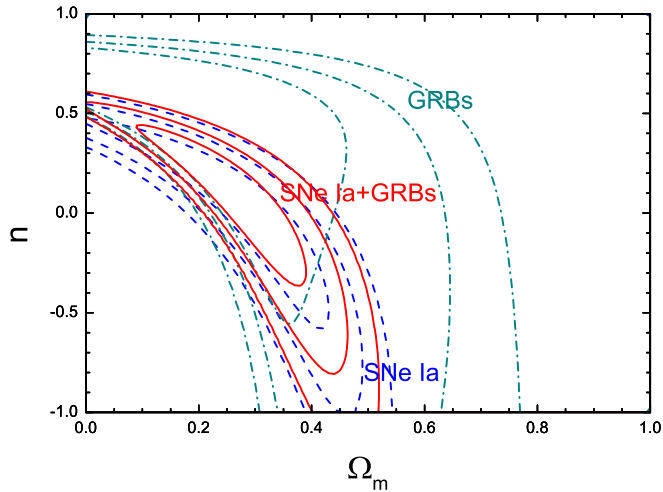}}
\caption{The contour confidence levels of $(\Omega_{m}, n)$ in the cardassian expansion model
   from the data for 82 GRBs ($z>1.4$) (dark cyan dash-dotted lines), 580 SNe Ia (blue dashed lines),
   and 82 GRBs $+$ 580 SNe Ia (red solid lines), respectively. The contours correspond to 1, 2, and $3\sigma$
   confidence regions. \label{Cardassian}}
\end{center}
\end{figure}

Fig.~\ref{Cardassian} shows constraints on $\Omega_{m}$ and $n$ from $1\sigma$ to
$3\sigma$ confidence regions by fitting observational data. The dark cyan dash-dotted lines and blue
dashed lines represent the results from 82 GRBs and 580 SNe Ia, respectively. The red
solid contours show the constraints from the combination of these data. The best values are
$\Omega_{m}=0.24_{-0.15}^{+0.15}$ and $n=0.16_{-0.52}^{+0.30}$ at the $1\sigma$ confidence level
with $\chi_{\rm min}^{2}=727.31$ for 659 degrees of freedom. This result is consistent
with the $\Lambda$CDM cosmology $(n=0)$ in the $1\sigma$ confidence region.

\subsection{$w(z)=w_{0}$ Model: Constant Equation of State}
\begin{figure}[htb]
\begin{center}
\resizebox{10.0cm}{!}{\includegraphics{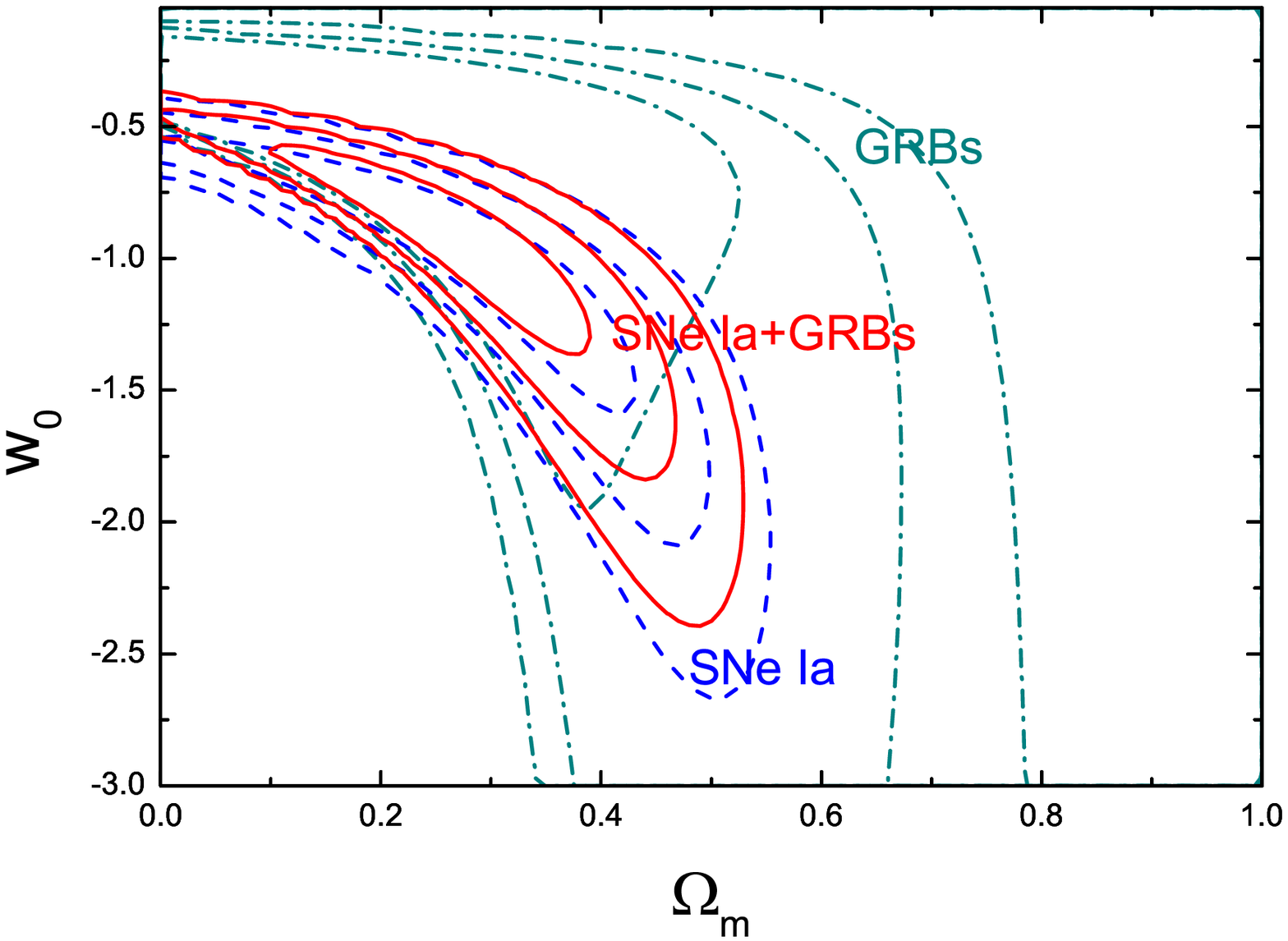}}
\caption{Constraints on $\Omega_{m}$ and $w_{0}$ from $1\sigma$ to $3\sigma$ confidence regions
   with dark energy whose equation state is constant. The contours are derived from GRBs
   (dark cyan dash-dotted lines), SNe Ia (blue dashed lines), and SNe Ia $+$ GRBs (red solid lines),
   respectively. \label{dark}}
\end{center}
\end{figure}

For the dark energy model with a constant equation of state $(w(z)=w_{0})$,
the luminosity distance for a flat universe is \cite{Riess04}
\begin{equation}
D_{L}(z)=cH_{0}^{-1}(1+z)\int_{0}^{z}dz[(1+z)^{3}\Omega_{m}+(1-\Omega_{m})(1+z)^{3(1+w_{0})}]^{-1/2},
\end{equation}
then the likelihood function depends on $\Omega_{m}$ and $w_{0}$. Fig.~\ref{dark}
shows the likelihood contours on $(\Omega_{m}, w_{0})$ plane for GRBs (dark cyan
dash-dotted lines), SNe Ia (blue dashed lines), and SNe Ia + GRBs (red solid lines),
respectively. The contours correspond to 1, 2, and $3\sigma$ confidence regions,
respectively. The cosmological parameters with the largest likelihood are $\Omega_{m}
=0.24_{-0.14}^{+0.16}$ and $w_{0}=-0.85_{-0.51}^{+0.28}$ ($1\sigma$) with
$\chi_{\rm min}^{2}=727.32$ for 659 degrees of freedom. For a prior of $\Omega_{m}=0.29$,
we obtain $w_{0}=-0.95_{-0.18}^{+0.14}$, which is consistent with the cosmological
constant (i.e., $w_{0}=-1$) in a $1\sigma$ confidence region.

\subsection{$w(z)=w_{0}+w_{1}z/(1+z)$ model: Time-dependent Equation of State}

We next examine models in which dark-energy changes with time. As shown above,
we adopt a simple model in which the dark-energy equation of state can be parameterized
by \cite{Chevallier01,Linder03}
\begin{equation}
w(z)=w_{0}+w_{1}z/(1+z)=w_{0}+w_{1}(1-a).
\end{equation}
The $\Lambda$CDM model is recovered when $w_{0}=-1$ and $w_{1}=0$. In this dark-energy
model, the luminosity distance is calculated by
\begin{equation}
D_{L}(z)=cH_{0}^{-1}(1+z)\int_{0}^{z}dz[(1+z)^{3}\Omega_{m}+(1-\Omega_{m})(1+z)^{3(1+w_{0}+w_{1})}e^{-3w_{1}z/(1+z)}]^{-1/2}.
\end{equation}

\begin{figure}[htb]
\begin{center}
 \resizebox{10.0cm}{!}{\includegraphics{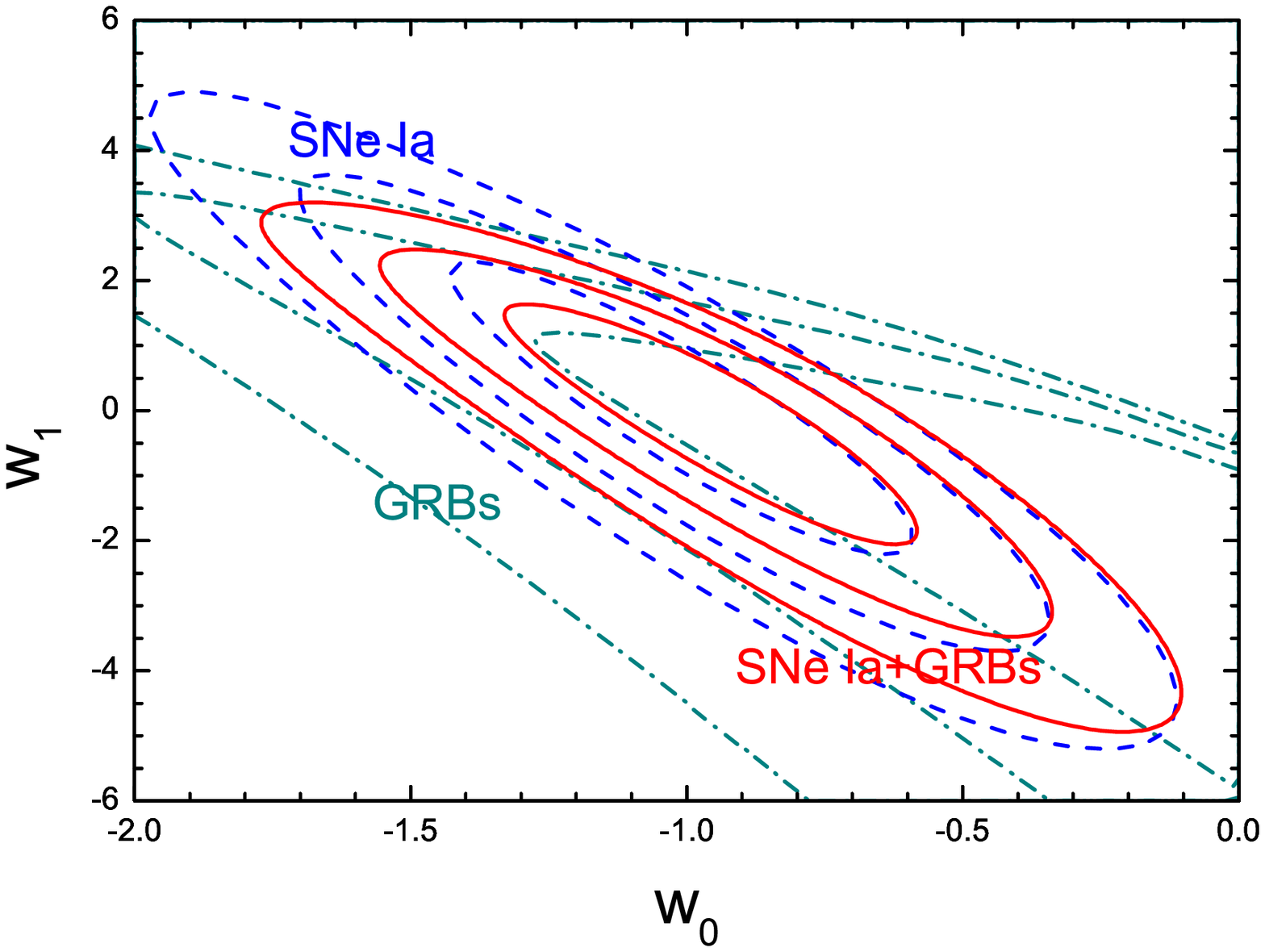}}
\caption{Constraints on $w_{0}$ and $w_{1}$ from $1\sigma$ to $3\sigma$ confidence regions
   with dark energy whose equation state is $w(z)=w_{0}+w_{1}z/(1+z)$. The contours are derived from GRBs
   (dark cyan dash-dotted lines), SNe Ia (blue dashed lines), and SNe Ia $+$ GRBs (red solid lines),
   respectively. \label{doubledark}}
\end{center}
\end{figure}

Fig.~\ref{doubledark} shows the constraints on $w_{0}$ versus $w_{1}$ from $1\sigma$ to
$3\sigma$ confidence regions. The dark cyan dash-dotted lines and blue dashed lines
represent the constraints from 82 GRBs and 580 SNe Ia, respectively. The red solid contours
are obtained from the combination of these data. For a prior of $\Omega_{m}=0.29$,
we find the best dark-energy parameters set is $(w_{0}, w_{1})=(-0.96_{-0.36}^{+0.39},
-0.04_{-1.96}^{+1.72})$ at the $1\sigma$ confidence level with $\chi_{\rm min}^{2}=727.54/659$.
This result is also consistent with the $\Lambda$CDM model (i.e., $w_{0}=-1$ and $w_{1}=0$)
in the $1\sigma$ confidence region.

\section{Conclusions and Discussion}\label{conclusions}

In this paper, we have updated five GRB luminosity relations $(\tau_{\rm lag}-L,
E_{\rm p}-E_{\rm iso}, E_{\rm p}-L, E_{\rm p}-E_{\rm \gamma}, \tau_{\rm RT}-L)$ among
certain spectral and light-curve features with the latest 139 GRBs. We find that the
five relations indeed exist with the latest GRBs data. To avoid any assumption
on cosmological models, we obtained the distance moduli of 57 low-\emph{z} ($z<1.4$) GRBs
by using cubic spline interpolation from the 580 Union2.1 SNe Ia compiled in Ref.~\cite{Suzuki12}.
Then, we calibrated the five relations with these 57 low-\emph{z} GRBs. In order to
constrain cosmological models, we extended the five calibrated luminosity relations to
high-\emph{z} and derived the distance moduli of 82 high-\emph{z} ($z>1.4$) GRBs.

Motivated by these significant updates of the observational data, we considered the joint
constrains on the Cardassian expansion model and dark energy with 580 Union2.1 SNe Ia
sample ($z<1.4$) and 82 calibrated GRBs data ($1.4<z\leq8.2$). In the $\Lambda$CDM cosmology, we find that adding
82 high-\emph{z} GRBs to 580 SNe Ia significantly improves the constrain on $\Omega_{m}-
\Omega_{\Lambda}$ plane. We obtain $\Omega_{m}=0.27_{-0.06}^{+0.08}$ and $\Omega_{\Lambda}
=0.62_{-0.19}^{+0.18}$ $(1\sigma)$. For a flat Universe, the contours yield $(\Omega_{m},
\Omega_{\Lambda})=(0.29_{-0.04}^{+0.04}, 0.71_{-0.04}^{+0.04})$. The transition redshift
at which the Universe switched from deceleration to acceleration phase is $z_{T}=0.64_{-0.14}^{+0.08}$
$(1\sigma)$. In the Cardassian expansion model, we obtain $\Omega_{m}=0.24_{-0.15}^{+0.15}$
and $n=0.16_{-0.52}^{+0.30}$ $(1\sigma)$. This result is consistent with the $\Lambda$CDM
cosmology $(n=0)$ in the $1\sigma$ confidence region. We also fit two dark energy models,
including the flat constant $w$ model (i.e., $w(z)=w_{0}$) and the time-dependent $w$ model
(i.e., $w(z)=w_{0}+w_{1}z/(1+z)$). Based on our analysis, it can be seen that our Universe
at higher redshift up to $z=8.2$ is consistent with the concordance model $(\Omega_{m}=0.27,
\Omega_{\Lambda}=0.73, w_{0}=-1, w_{1}=0)$ within $1\sigma$ level. These results suggest that
time dependence of the dark energy is small even if it exists.

Since the discoveries of distance indicators of GRBs, these luminosity indicators have been used
as standard candles for cosmological research at high redshifts. However, the dispersion of
distance indicators are still large, which restricted the precision of distance measurement
by GRBs. The large dispersion may be due to that some contamination of the GRB sample is
unavoidable, and that pure luminosity indicators may never be found for these sources. Of course,
it could also due to that we simply have not yet identified the correct spectral and lightcurve
features to use for these luminosity relations. On the other hand, it could also due to that
we are inevitably suffering from the systematic errors and intrinsic scatter associated
with the data. In order to estimate distance of GRBs more precisely, we should take
efforts to investigate possible origins of dispersion of the distance indicators, and/or
search for more precise distance indicators in the future.

\vspace*{0.5cm}

\section{Acknowledgements}

This work is partially supported by the National Basic Research Program (``973" Program)
of China (Grants 2014CB845800 and 2013CB834900), the National Natural Science Foundation
of China (grants Nos. 11322328 and 11373068), the One-Hundred-Talents Program,
the Youth Innovation Promotion Association, and the Strategic Priority Research Program
``The Emergence of Cosmological Structures" (Grant No. XDB09000000) of
the Chinese Academy of Sciences, and the Natural Science Foundation of Jiangsu Province.

\end{document}